\documentclass[runningheads,a4paper]{llncs}
\usepackage{amssymb}
\usepackage{graphicx}

\usepackage{url}
\urldef{\mailsa}\path|{prasenjitk, gopi}@csa.iisc.ernet.in|

\usepackage{algpseudocode}
\usepackage{comment}
\usepackage{amsmath}
\usepackage{amssymb}
\usepackage{enumitem}
\usepackage{graphicx}
\usepackage[T1]{fontenc}
\usepackage{float}
\usepackage{amstext}
\usepackage{amssymb}
\usepackage{wrapfig}
\usepackage{times}
\usepackage{verbatim}
\usepackage{textcomp} 
\usepackage{multirow}
\usepackage{morefloats}
\usepackage{algorithmicx}

\newtheorem{defi}{Definition}
\usepackage{epsfig}

\floatstyle{ruled}
\newfloat{algorithm}{tbp}{loa}
\floatname{algorithm}{Algorithm}
\newfloat{equation}{tbp}{loa}
\floatname{equation}{Equation}
\usepackage{subfigure}

    \def\independenT#1#2{\mathrel{\setbox0\hbox{$#1#2$}%
    \copy0\kern-\wd0\mkern4mu\box0}} 


\makeatother
\begin{document}
\mainmatter  

\title{Scalable Reliability Modelling of RAID Storage Subsystems}


%
%
\author{Prasenjit Karmakar%
\and K. Gopinath}
%

\institute{IISC Bangalore}

%
%

\toctitle{Lecture Notes in Computer Science}
\tocauthor{Authors' Instructions}
\maketitle

\subsection*{Abstract}
Reliability modelling of RAID storage systems with its various
components such as RAID controllers, enclosures, expanders,
interconnects and disks is important from a storage system designer's
point of view. A model that can express all the failure
characteristics of the whole RAID storage system can be used to
evaluate design choices, perform cost reliability trade-offs and
conduct sensitivity analyses. However, including such details makes
the computational models of reliability quickly infeasible.

We present a CTMC reliability model for RAID storage systems that scales to
much larger systems than heretofore reported and we try to model all
the components as accurately as possible. We use several state-space
reduction techniques at the user level, such as aggregating all in-series components and
hierarchical decomposition, to reduce the size of our model. To
automate computation of reliability, we use the PRISM model checker as
a CTMC solver where appropriate. We use both variations of PMC $-$ numerical as well as statistical model checking 
according to the size of our model. Our modelling
techniques using PRISM are more practical (in both time and effort)
compared to previously reported Monte-Carlo simulation techniques.


Our model for RAID storage systems (that includes, for example,
disks, expanders, enclosures) uses Weibull distributions for disks and, where
appropriate, correlated failure modes for disks, while we use exponential
distributions with independent failure modes for all other components.
To use the CTMC solver, we approximate the Weibull distribution for a
disk using sum of exponentials and we confirm that this model gives
results that are in reasonably good agreement with those from the
sequential Monte Carlo simulation methods for RAID disk subsystems
reported in literature earlier.
Using a combination of scalable techniques, we are able to model and compute
reliability for fairly
large configurations with upto 600 disks using this model. 
\section{Introduction}
\label{sec1}
Despite major efforts,  both in industry and in academia,  achieving
high reliability remains a major challenge in large-scale IT
systems. A particularly big concern is the reliability of storage
systems because failure can not only cause temporary data
unavailability but also to permanent data loss in the worst case.

The reliability of RAID storage systems used in data centres
or in critical server applications needs to be high. These systems consist of many components such as RAID
controllers, enclosures, expanders, interconnects and, of course,
disks. Failures in any of these components can lead to downtime or
data loss, or both. Hence, redundancy is provided through, for example,
dual controllers and dual expanders for 
high availability. A multiplicative set of paths thus needs to be
considered for modelling reliability, but decomposition techniques
that have been 
reported recently\cite{trivedi-cloud} cannot handle the explosive
growth in the paths due to their reliance on locality and
fixpoint iterations across local modules.

While several studies have been conducted on understanding and
modelling ``disk'' failures \cite{disk1, disk3}, there seems to be little
work done on scalable analysis of the reliability of a ``whole'' RAID storage
system with its various components. 
Jiang et al. \cite{netapp_fast} presented an empirical analysis of
NetApp AutoSupport logs collected from about 39,000 storage systems
commercially deployed at various customer sites. 
An important finding of the study
is that component failures other than disks (such as those of
controllers, enclosures, SAS cables) contribute most
(27-68\%) to the failure of storage subsystems.
Another recent study \cite{osdi_10} by Ford et al. characterizes the end to end data availability properties of
cloud storage systems at the distributed file 
system level based on an extensive one year study of Google's main storage 
infrastructure. 
A key point is the importance of
modelling correlated failures when predicting availability, and show
their impact under a variety of replication schemes and placement policies.

To design and build a reliable RAID storage system, it is 
important to have a model of the storage system that expresses all the
storage failure characteristics of all of its components using which
we can do several ``what-if'' analyses for the storage system. Simulation (for example, written in plain C code) is a widely used and powerful technique for
evaluating such systems. Simulation is the most flexible
since it allows us to use arbitrary distributions (such as
Weibull common in reliability studies) and even traces. However, 
writing simulation code for such non-trivial systems is error-prone or the results cannot be validated easily.

To compute the reliability of a RAID storage system,  we need 
to model all the RAID components while keeping scalability in mind, but
previous work, to the best of our knowledge, has not considered such models.
In this work, we model all the components of a RAID storage system (controllers, enclosures, expanders, interconnects and disks)
with a failure rate and repair rate, and use scalable techniques for
computing reliability. In addition, we use a CTMC solver
from the PRISM verification tool\cite{PRISM} that is well known for its
computational efficiency and ability to minimize state spaces (it can
therefore handle in excess
of 100's of millions of states), provide a
powerful language that can easily express the desired reliability
queries and also provides results using simulation that samples paths on the same reliability model
when the state space becomes too large.

We assume exponential failure distribution for all the components in
the system except disks. For disks, we initially assume a simple
3-state model \cite{infant_mortality} and later we use a Weibull
model \cite{disk3}.  We approximate Weibull model using a sum of
exponentials and show that this model gives almost the same results as
the sequential Monte-Carlo simulation methods for disk
subsystems reported earlier \cite{disk3}.

However, the results of using this detailed Weibull model do not agree 
well with the field data for the RAID configurations we use for validation.
Hence we infer correlated failures in such configurations
and therefore revise our models by estimating and including correlated failures.
Since such models are computationally difficult, we use a variety of
techniques for scalability (such as
hierarchical decomposition); we are then able to model large RAID
configurations with upto 600 disks. The primary goal in this paper is
the development of the methodology for computing the reliability of
reasonable size storage systems while being as close as possible to real 
storage systems and validated against whatever field data is available.
As we do not have access to some parameters that are critical (for eg,
for correlated disk failures), we hope that such data will be increasingly
be collected and made available to researchers in the future.

Our models can be used to perform sensitivity analysis and make
cost-reliability tradeoffs. It can also help in throwing some light on
the suitability of Markov models for computing storage reliability.
Greenan \cite{Greenan} has suggested that only simulation can be
used to model rebuild progress but not Markov models due to their
memorylessness. However, we show that with the right models, it is
possible to simulate memory even with Markov models and we do get
similar results.

The outline of the paper is as follows:     
Section \ref{sec2} briefly describes the components present in storage systems
and presents some configurations. Section \ref{sec3} lists
definitions, model input parameters and modelling assumptions. Section \ref{sec4} shows the modelling of RAID storage systems using a 
simple disk model. Section \ref{sec5} 
shows detailed modelling of RAID disk subsystems assuming Weibull model and correlated failure. This section also discusses
validation of our model against field data. Section \ref{sec6} presents conclusions.
 \section{Storage System Architecture}
\label{sec2}
\subsection {Main Components in RAID Storage Systems}
\label{sec2.1}

\textbf{RAID Controllers}: These are usually a pair of
    controllers, one acting as the primary and the other acting as the backup. 
        For load balancing,  one controller is the primary for
        half the disks in the system that also acts as the secondary for the rest of
        the disks; it is vice versa for the other controller.

\textbf{Enclosures}: All the disks reside in a
  component called ``external storage enclosure'' for expandability
  and portability. 
       In each enclosure,  there are several components such as
       redundant power supply/cooling fans and midplane for all the disks. 
      An enclosure fails if both the power supplies fail or
      both the cooling fans fail,  or the midplane fails.
      As enclosure
      components are shared by all the disks inside, an  
      enclosure's reliability depends on the number of disks inside.  

\textbf{Expanders}: This is a fan-out switch used in large
  storage systems to connect multiple initiators and targets for
  scalability and fault-tolerant path redundancy.

\textbf{Interconnects} are usually SAS (Serially Attached Storage)
cables for connecting system components.


\textbf{Disks} in the system are SAS or SATA disks. SAS controllers support both SAS and SATA disks.
\subsection{Some Storage System Configurations}
\label{sec2.2}
\begin{figure}[htp]
  \centering
  \subfigure[1 enclosure]{\label{fig1:a}\includegraphics[width=.49\linewidth]{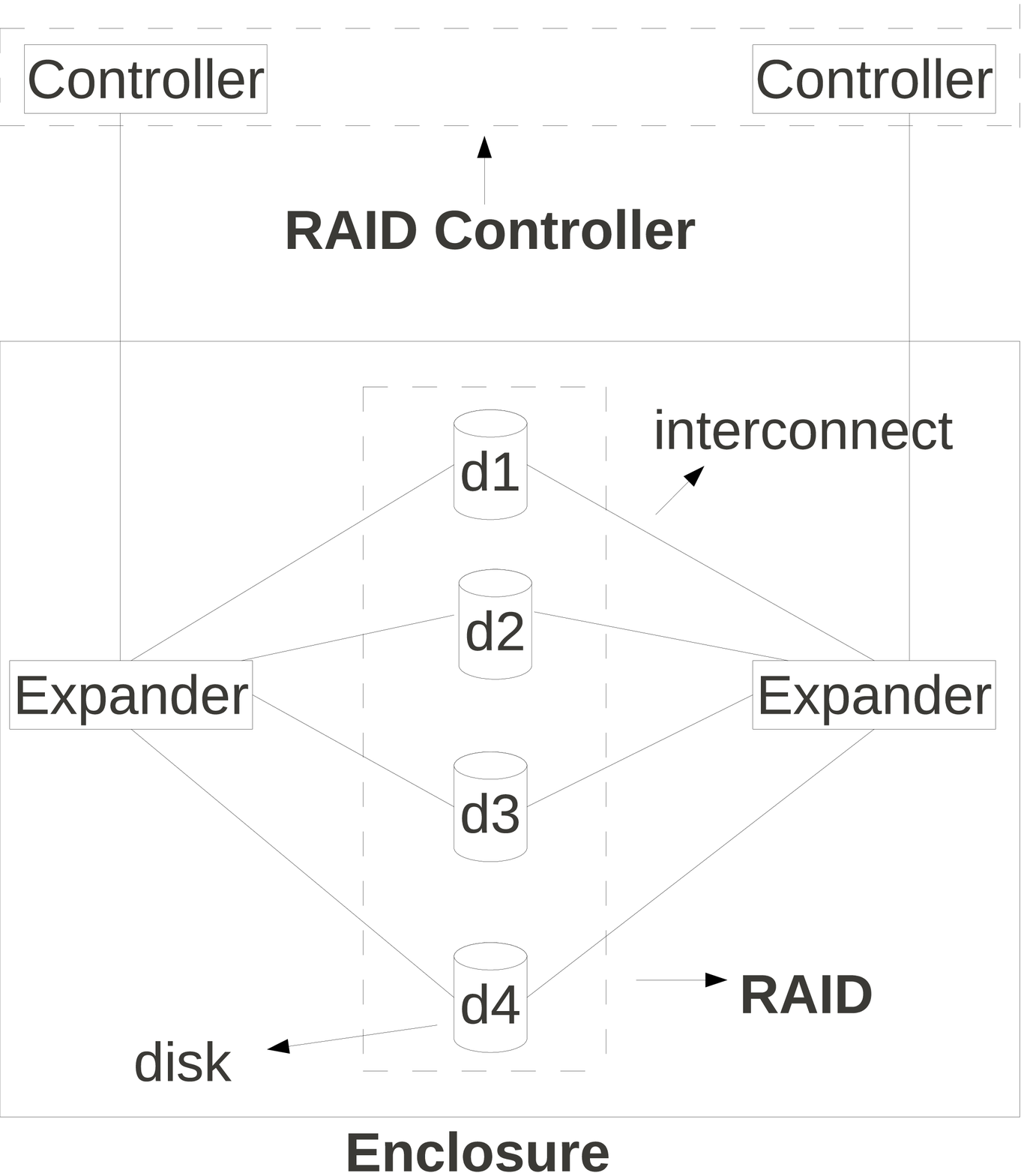}}\hfill
  \subfigure[2 enclosure]{\label{fig1:b}\includegraphics[width=.49\linewidth]{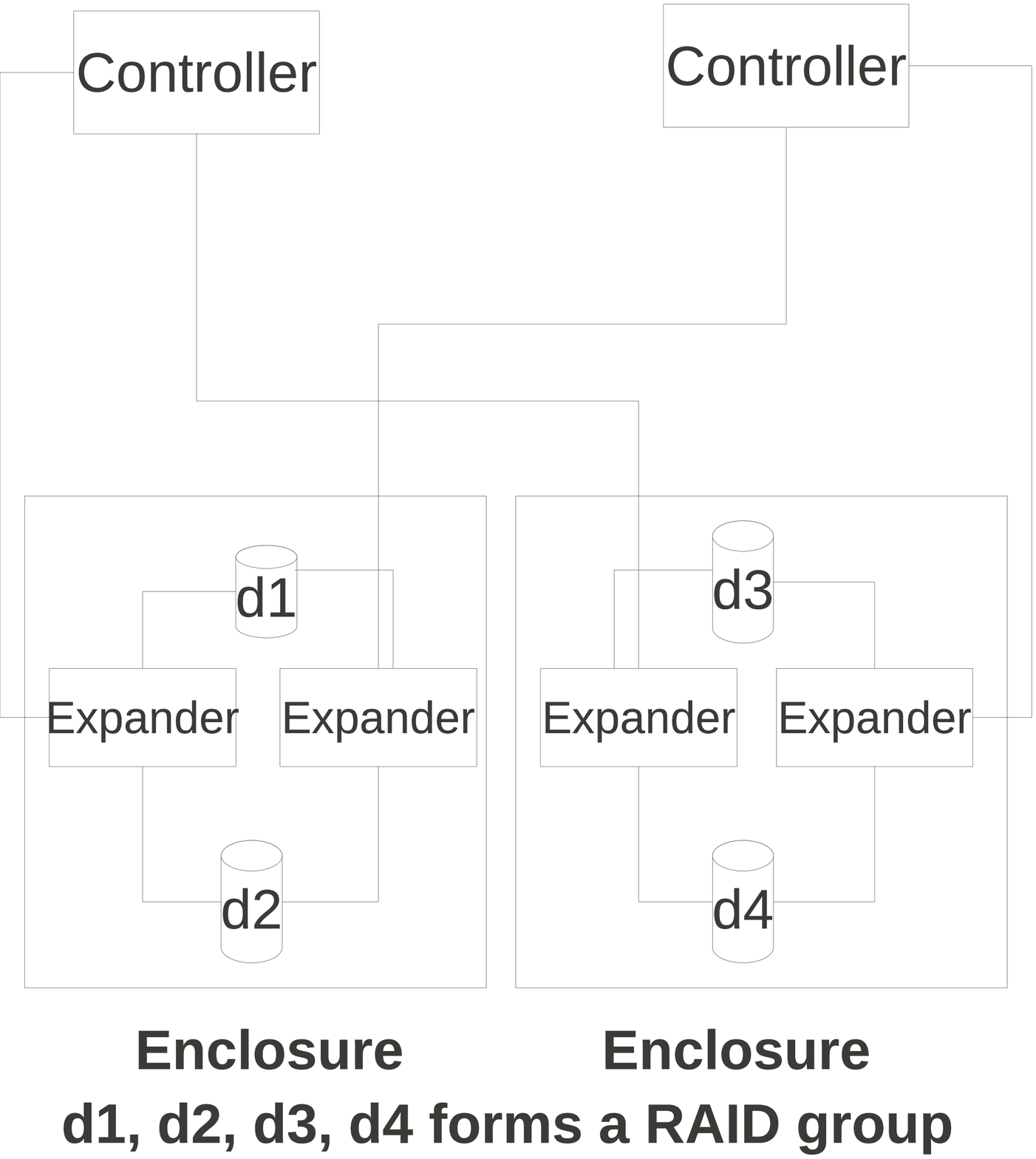}}
\caption{4 disk RAID5 in 1 and 2 enclosures}
\label{fig1}
\end{figure}

Figs. \ref{fig1:a} and \ref{fig1:b} show a 4 disk RAID5 group
in one enclosure and across 2 enclosures respectively. There are multiple types of
``redundancy'' present:
\begin{enumerate}
 \item redundant controllers, interconnects, expanders
 \item redundant disks (RAID)
 \item redundant enclosures (with ``spanned'' RAID groups
   across)
\end{enumerate}
While the first two redundancy mechanisms clearly increase the reliability of
the system, analysis is needed to know when spanning is beneficial. 
\section{Definitions and Assumptions} 
\label{sec3}
\begin{defi}
 \textbf{[RAID5 Reliability]}
A RAID5 group experiences ``data inaccessibility or data loss'' (DIL) if
  \begin {enumerate}
   \item data in any of two disks in the RAID5  group are
     inaccessible. 
     The data in a disk is said to be inaccessible 
if some component in its access path fails or the disk itself fails.
   \item Or, a data of a disk is inaccessible and an unrecoverable
     error occurs during rebuild of the data.
  \end {enumerate}
\end{defi}
We can extend the above definition to RAID6 and other RAID systems by
changing the number of 
inaccessible data disks in the first part of the definition.
\subsection{Newer Reliability Measures}
\label{sec3.1}
\begin{defi}
 \textbf{[MTTDIL]}
        Mean Time before ``any'' of the RAID groups experiences ``data
        inaccessibility or data loss'' (DIL). 
This denotes the average time before which at least one user using the system experiences data unavailability. 
\end{defi}

Note that MTTDIL has lower value than MTTF as inaccessibility of data is also
taken into account, not just data loss. It is therefore a composite
measure of availability and reliability.
Given the computational flexibility with the use of PRISM
toolset \cite{PRISM}, it is possible to compute specific
reliability measures as needed by the system at hand. Due to lack of space we describe them in 
Appendix \ref{app1}

We have chosen the ``mean'' as our reliability measure even though it
may not be a good reliability metric  \cite{markov_bad} as the field
data available with us provide only the ``mean'' values. 
\subsection{Modelling assumptions}
\label{sec3.2}
Due to the lack of relevant disk failure parameters in the field data
(as needed in detailed disk models such as \ref{sec5.1}), we build our
models incrementally and, at each stage, check with field data available for the
storage system as a whole.  
\begin {enumerate}
 \item Initially, we assume uncorrelated failures across components.
Later,  we consider correlated failures for disks in our model
       and show that the model results match the field data available
       to us. 
 \item For a disk, we assume at the start a simple 3-state Markov model (with 
       burn-in rate,  pre-burn-in failure rate,  post burn-in failure rate \cite{infant_mortality}) but
       later consider Weibull models. We assume constant failure rate
       for all other components as we have access only to MTTF values.
 \item We also assume constant repair rate for all the components but
   this is not necessary.
\end {enumerate}
\subsection{Model input parameters}
\label{sec3.3}
Table \ref{table2} shows MTTF values of the components obtained from a
few storage
vendors\footnotemark[1]. Meanwhile, disk MTTF has been taken from previous 
literature \cite{disk1,disk_mttf}. We have used Mean Time to Repair (MTTR) of a
non-critical component as 30 min based on some inputs from industry.  
\footnotetext[1]{Some of the field data have been generously given to us by a
  storage vendor but requesting no attribution.}
\footnotetext[2]{The enclosure type we consider here can contain atmost 24 disks.}
\begin{table}[h]
\begin{center}
 
\begin{tabular}{|c|c|}
\hline
 \textbf{Components} & \textbf{MTTF value} \\
\hline
Disk                    &           33 yr       \\ \hline    
Controller		&	604440 hr	\\ \hline
Expander                &       2560000 hr	\\ \hline
Enclosure\footnotemark[2]		&	28400 hr if $\leqslant$ 50\% full, 
                                else 11100 hr  \\ \hline
Interconnect		&	200000 hr\\ \hline
\end{tabular}
\end{center}
\caption{MTTF of components}
\label{table2}
\end{table}

For all the computational results reported in this paper,  
we have used a 2.8GHz 8GB RAM machine with 16GB swap space.
\section{
Modelling RAID5 systems with a simple disk model
}
\label{sec4}
In the beginning, we give a brief introduction to other possible modelling aprroaches and compare them with our approach 
using PRISM. In the later subsections we describe our models using PRISM.
\begin{figure}[h]
\centering
\includegraphics[width=3in]{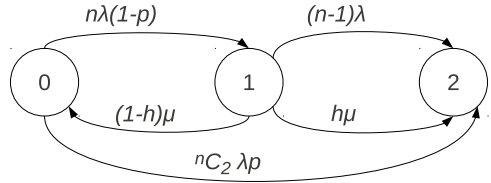}
\caption{Model of a $n$ disk RAID5 in an enclosure assuming disk correlated failure; 
State 0: working,  1: one disk fails,  2: data loss, $h$:
probability of unrecoverable error during rebuild, $\lambda$: disk
failure rate, $\mu$: rebuild rate}
\label{fig12}
\end{figure}   
\subsection{PRISM: Comparison with other approaches}
Continuous Time Markov Chains (CTMC) have been used widely to build
reliability models of systems. 
For example, 
a simple Markov reliability model of RAID5
is given by Rao et al. \cite{disk1} (see Fig.\ref{fig12} with $p$=0). For more complex configurations
or for modelling other components such as enclosures, tool-enabled
computational models are critical.

Another approach for evaluating such complex systems is Monte-Carlo simulation 
techniques where one has to manually find all the system failure cases by keeping a timeline for each of the components. However, 
writing simulation code for such non-trivial systems is error-prone or the results cannot be validated easily.

Here,  we use a model checking tool,  PRISM (Probabilistic Symbolic Model Checker) \cite{PRISM},  to build 
and analyze the CTMC models. 
PRISM has a modular
process algebra based language with a probability and reward operator, and it 
supports quantitative model checking. This is suitable for modelling
the reliability of systems as the failure mode of each component can be 
described in a module separately. 
Given a PRISM program, PRISM uses matrix computations for
exact model checking (by numerical solution technique) of logical formulae (in continuous stochastic
logic, CSL),  as
well as a sampling based simulation approach that is suitable when the
state size is very large. We use both approaches according to the size of the model.
All our models are reliability models and hence there
are no transitions in our models out of the ``data inaccessible or loss'' state of a
specific instance of a RAID system when we calculate MTTDIL.  
 


Our modelling environment in PRISM is much more feasible for modelling large complex storage systems compared to the Monte-Carlo simulation 
techniques 
as, in case of PRISM, the tool 
itself, in effect, does the work of finding all the system failure cases for us, by building a CTMC model of the whole system, from model description
 written in the PRISM modelling language. Moreover, in case of rare-event failures simulation may take a long time compared to the PRISM model checking 
(we show such an example in Section \ref{sec 5.1.1}).

In the following section,  we describe several modelling and state-space reductions techniques for our systems.
Section \ref{sec4.1} presents results of modelling of some small systems mainly using PRISM model checker. Section \ref{sec4.2} uses PRISM 
discrete-event simulator to calculate reliability measures for larger systems. Section \ref{sec4.3} presents the hierarchical decomposition
technique to model even larger systems. Section \ref{sec4.4} presents
results of modelling some known field configurations (such as multiple 
controller pair configurations) along with how they agree with field data. 
\subsection{PRISM model of small systems}
\label{sec4.1}
\begin{figure}
\centering
\includegraphics[height=2.0in]{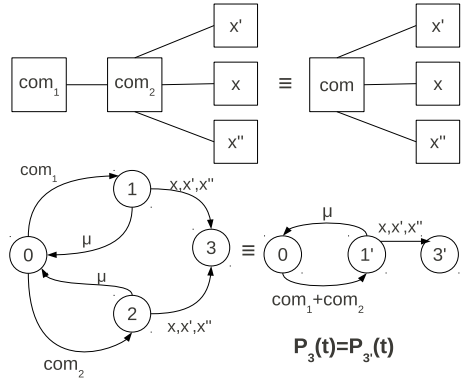}
\caption{Composing in-series components: $com_{1}$ and $com_{2}$ are in series and 
can be replaced by series equivalent $com$ as the Markov model at right is equivalent to the model at
left. $x$/$x'$/$x''$ stands for the transitions corresponding to other 
components i.e. those that are not connected in series with $com_{1}$ and $com_{2}$. State 1' is 
the merged equivalent of states 1 and 2. $P_{i}(t)$ is the transient probability of reaching state $i$}
\label{fig3}
\end{figure}

In RAID systems,  there are some components connected in
series (such as the controller,  interconnect and expander in Fig.\ref{fig1:a}). 
To reduce the state space,  we can compose them: if there is a set of components $com_{1}$,  $com_{2}$, ...., $com_{k}$ in series 
and if each component has the same repair rate,  then we can replace the set 
by an equivalent component $com$ such that
failure rate of $com$ = $\sum_{i=1}^k $ $f(i)$,  where $f(i)$ is failure rate of the $i$-th component in series and repair rate  of 
$com$ is $\mu$ where $\mu$ is repair rate of each component (Fig.\ref{fig3}).

\begin{table}[ht]
\begin{center}

\begin{tabular}{|c|l|c|c|c|c||c|l|c|}
\cline{4-9}
\multicolumn{3}{c|}{} & \multicolumn{3}{|c||}{\textbf{Enclosure MTTF indep. of no. of disks}} & \multicolumn{3}{|c|}{\textbf{Enclosure MTTF $\propto$ no. of disks}}\\
\hline
\textbf{$m$} & \textbf{Distribution of disks} & \textbf{States} & \textbf{$t$=12} & \textbf{Time(hr)} & \textbf{Model size(GB)} & \textbf{$t$=2} & \textbf{$t$=3} & \textbf{$t$=4}\\
\hline

1                    & $-$         &     14827521   	&   27697	   &   0.2  & 0.3 & 10975 & 10975 & 10975  \\ \hline
\multirow{4}{*}{2}  & (7+1)        &     20803073  	&   27676          &  0.5  & 0.4& 10971& 10971 & 10971 \\
                    & (6+2)         &     36399361     &   14004           &   1     &0.7      & 7911& 7911 & 7911  \\
                    & (5+3)         &     35961985    &   14004            &  1.1      &  0.8    & 5513 & 7911 & 7911 \\
                    & (4+4)        &    34561793 	&   14004	   &   0.6  & 0.8 & 5513&5513 &14004  \\ \hline
\multirow{5}{*}{3}  & (6+1+1)       &     36115713      &   27654	  &   1.2  & 0.7 &10967 &10967 &10967  \\
                    & (5+2+1)        &     77596289    &    13998              & 4.3 & 2.1   &  7909& 7909 &7909  \\   
                    & (4+3+1)        &     62729729    &   13998	   &   3  & 1.8 & 5512 & 7909  & 13998  \\
                    & (3+3+2)         &    114833281    &   9372           &  4.6 &7.5  &3682 & 9372 & 9372     \\   
                    & (4+2+2)         &     96482561   &    9372	   &   3.6  & 2.9& 6185& 6185& 9372  \\ \hline
\multirow{4}{*}{4}  & (5+1+1+1)       &     28689665    &   27630	  &   0.5  & 0.5 & 10962&10962 &10962  \\
                    & (4+2+1+1)        &     53671681  &   13992	   &   1.4 & 1 &7907 &7907 &13992  \\
                    & (3+1+2+2)       &    109296129  &    9369          &    4.6  & 2.1 & 6183&9369 &9369  \\
                    & (2+2+2+2)       &    145866497  &    7043          &   1.8   & 8.1 & 7043&7043 & 7043 \\ \hline
\multirow{3}{*}{5}  & (4+1+1+1+1)      &     42528513   &   27606	  &   1.8  & 0.8 &10958 & 10958 & 27606 \\
                    & (3+1+2+1+1)      &     66852353   &   13986        &    4  & 1.5  &7904 &13986 & 13986 \\
                    & (2+2+2+1+1)       &    100492801 &    9366          &   3.6  & 1.9  & 9366 & 9366 & 9366  \\ \hline 
\multirow{2}{*}{6}  & (3+1+1+1+1+1)    &     44301697 &   27581           &   0.9  & 0.7  &10954 & 27581  &  27581  \\
                    & (2+2+1+1+1+1)     &     67861761 &   13979	   &   1  & 1.0  & 13979& 13979& 13979 \\ \hline
7                   & (2+1+1+1+1+1+1)  &     45208577 &   27360	    &   1  & 0.6  &27360 &27360 & 27360\\ \hline 
8 & (1+1+1+1+1+1+1+1)           & 8649402441 &  \textbf{OOM} & $-$ &  $-$  &  $-$ &  $-$  &  $-$\\ \hline

\end{tabular}

\end{center}
\caption{MTTDIL (hr) of a 8 disk RAID5; $m$: number of enclosures; $t$ is the threshold number of disks in an enclosure (max 24 disks) after which its MTTF decreases; \textbf{OOM}: Out of Memory Error; 
indep.: independent; $\propto$: dependent. Note that MTTDIL of 7000
hours is not indicative of unacceptable data loss as it incorporates
non-availability due to interconnect failure.}
\label{table3}
\end{table}
We replace all the final states (corresponding to different types of
failures such as those of
enclosure,  expander or  disk) in our model by a single ``fail state''. Using 
these two optimizations,  we are able to reduce the model size and
execution time significantly (although, execution time still mainly depends on model input parameters). 

Table \ref{table3} shows MTTDIL of a 8 disk RAID5 with $m$ enclosures ($m\leqslant8$) calculated in PRISM. The second column 
denotes the distribution of disks across enclosures. For $m$ enclosures,  $(i_1+i_2+......+i_m)$ 
denotes a configuration with $m$ enclosures where enclosure $j$ contains $i_j$ number of disks of the RAID group. Table \ref{table3} also shows the MTTDIL values assuming variable failure rate for an enclosure i.e.

{\it enclosure MTTF} = 28400 hr if no. of disks in it $\leqslant$ $t$
              
~~~~~~~~~~~~~~~~~~~~~~~~~~~~= 11100 hr otherwise

We have generalized the value of $t$ from 2 to 4 (while for field data it
is 50\% occupancy) to understand its impact. Given $m$
enclosures, each with some capacity $c$, it is not possible to say
what the optimal configuration is without detailed modelling. 

The important findings from the analysis of PRISM models are:
\begin{enumerate}
 \item A 10\% increase of enclosure
  MTTF causes MTTDIL to increase by 9.9\%. Hence, the enclosure 
  is the main determining component in the reliability of a RAID group.
 \item MTTDIL depends on the number of enclosures present in the system and the distribution of disks of a RAID group across enclosures. 
For example,  consider three cases (Table \ref{table3}) of
distribution of disks in 2 enclosures (4+4),  3 enclosures (6+1+1) and 4 enclosures (3+1+2+2).
   In the first case,  if ``any'' of the enclosures fail ``data inaccessibility'' occurs while in the second case only failure of enclosure
1  causes ``data inaccessibility''. Hence,  spanning increases reliability here. In the third case,  out of 4 enclosures, failure of 
 3 enclosures causes ``data inaccessibility''. Hence,  spanning decreases reliability
compared to case 1. 
\end{enumerate}

Based on the results of Table \ref{table3} we have designed a algorithm for spanning a RAID group across enclosures which is described in
detail in Appendix \ref{app2}.



 
\begin{table}[ht]
\begin{center}
\begin{tabular}{|c|c|c|c|c|}
\hline
 \textbf{Configs.} & \textbf{MP} & \textbf{SP}  & \textbf{Gain} & \textbf{Extra Cost}\\
\hline
(1, 2, R1)	&  28400	&      23665	 & 1.2 & (1, 1, 3) \\ \hline
(2, 2, R1)	& 6.68E8	&      3.39E8	 & 1.97 & (0, 2, 4)\\ \hline
(2, 2, R1)	& 6.68E8	&      603561	 & 1106 & (1, 2, 4)\\ \hline
(1, 4, R5)	&  28245	&       23552	 & 1.2 & (1, 1, 5)\\ \hline
(2, 4, R5)	&  14157	&       11801	 & 1.2 & (0, 2, 6)\\ \hline
(2, 4, R5)	&  14157	&       12036	 & 1.18 & (1, 2, 6)\\ \hline
(4, 4, R5)        &  4629364        &      281707    & 16   & (0, 4, 8)\\ \hline
(4, 4, R5)        &  4629364        &      528003    & 8.7   & (1, 4, 8)\\ \hline
\end{tabular}
\end{center}
\caption{
Cost-reliability trade-offs: 
Configs. $(a, b, c)$: $a$: no. of enclosures,  $b$: no. of disks,  $c$: RAID group type (R1: RAID1, R5: RAID5);
 MP: multi-pathing and SP: single-pathing; Gain: reliability gain
 using multi-pathing; Extra cost $(a, b, c)$ due to multi-pathing: $a$: no. of controllers,  $b$: no. of expanders,  $c$: no. of SAS cables.
}   
\label{table4}
\end{table}
Jiang et al.  \cite{netapp_fast} showed that multi-pathing increases storage reliability. Table \ref{table4}  
shows some configurations both with multi-pathing and single-pathing,  and the corresponding MTTDIL values. 
 In some cases,  multi-pathing increases reliability by a factor of more than 1000 whereas in 
some other cases it is much lower. Based on such calculations,  cost-reliability trade-offs can be attempted. 
The redundancy of a component is
beneficial only if a single failure of some other component does not
cause ``data inaccessibility'' of a whole RAID group.
\subsection{Discrete Event Simulation}
\label{sec4.2}
Table \ref{table3} shows (for $m$=8) that it is not possible to model large systems consisting of multiple
RAID groups using PRISM model checker (due to state space explosion). 
To calculate the reliability measures for larger systems we use PRISM discrete-event simulator. 
This simulator generates a large number of random paths using the PRISM language model description
(without explicitly constructing the corresponding Markov Model), evaluates the result of the given
properties on each run, and uses this information to generate an approximately correct result. Currently, PRISM
simulator has support only for exponential distribution. 
We use a confidence parameter of 0.01 and 
``maximum path length'' of 1E9 to calculate our reliability measures. PRISM imposes a maximum path length to avoid the need to generate excessively long or
infinite paths.
\begin{figure}[htp]
  \centering
  \subfigure[single-pathing in encl. 3/4]{\label{fig5:a}\includegraphics[height=2.5in]{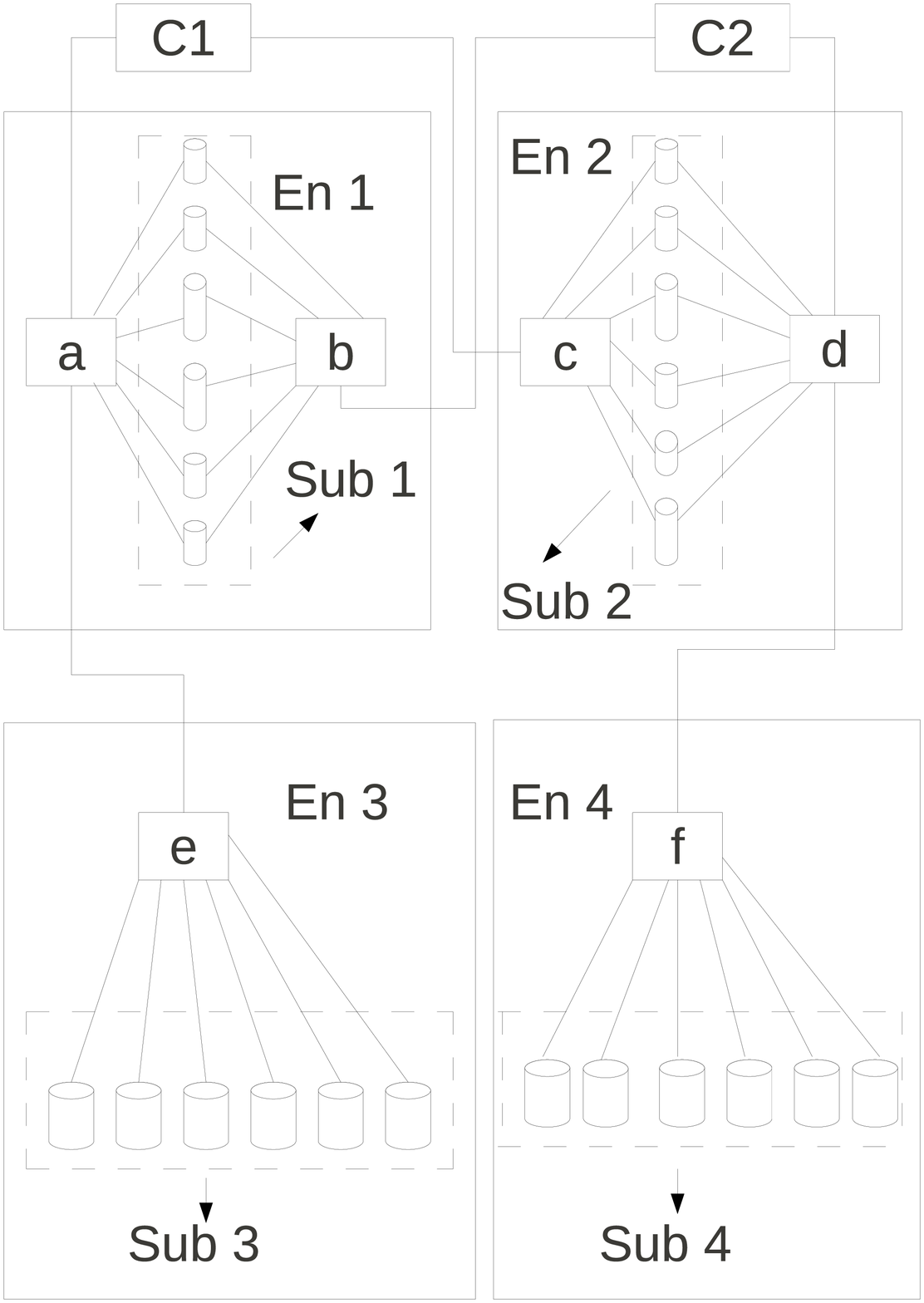}}\hfill
  \subfigure[multi-pathing]{\label{fig5:b}\includegraphics[height=2.5in]{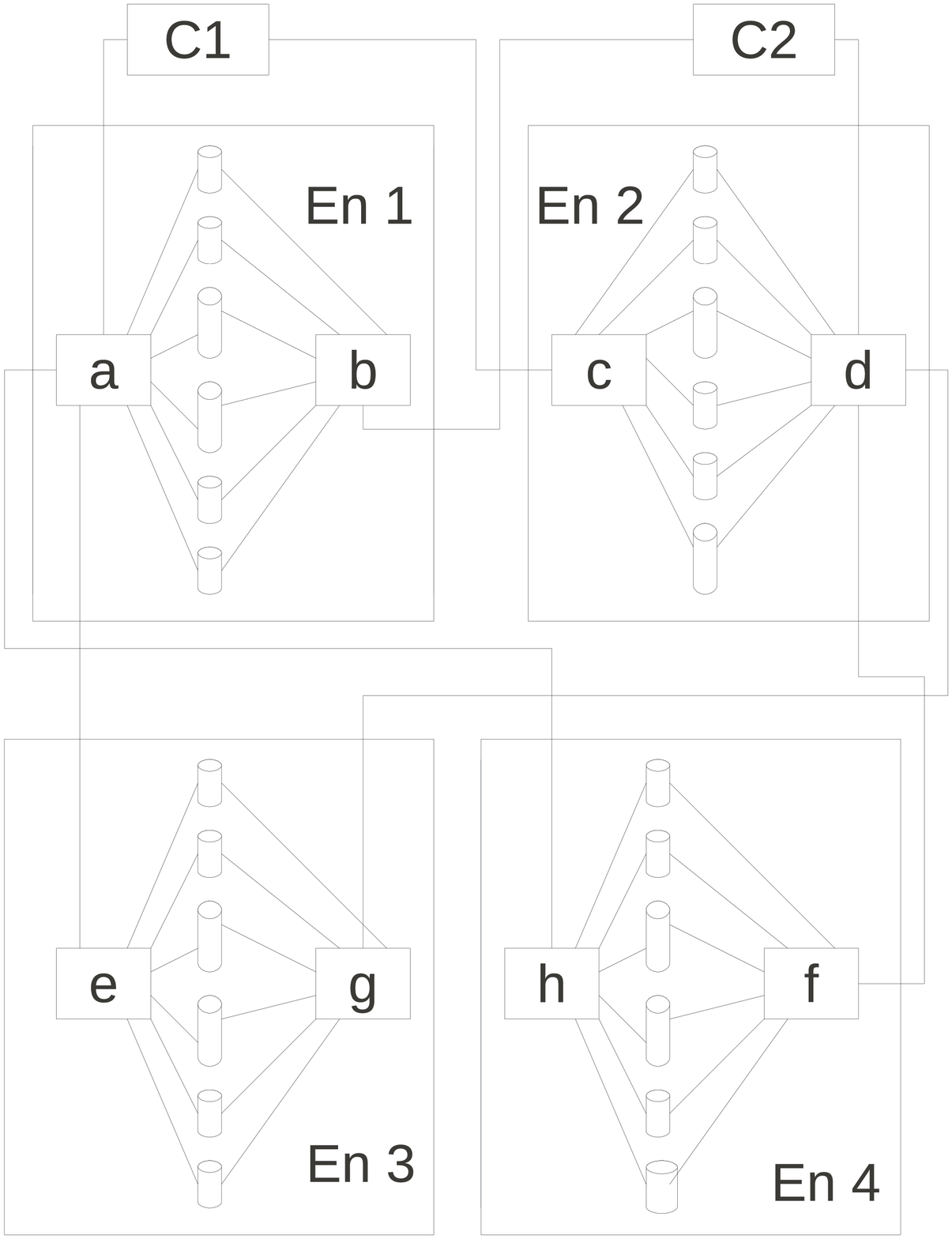}}
\caption{4 enclosures 24 disks as 4 RAID5 groups}
\label{fig5}
\end{figure}

Fig.\ref{fig5} shows two configurations with multiple RAID groups. 
Fig.\ref{fig5:a} shows a 4 enclosure 24 disk system with 4 RAID5 groups without redundant
paths for disks in enclosures 3 and 4 whereas in Fig.\ref{fig5:b}
redundant paths are present for disks in enclosures 3 and 4. 
Enclosures 3 and 4
are daisy-chained to enclosures 1 and 2 respectively as the controller may not 
have enough ports to connect all the enclosures present in the system directly.
Table \ref{table51} gives the MTTDIL
for these systems. 




It is not possible to simulate larger systems with the required confidence 
because the time for simulation increases rapidly and we need to decompose our models. 

\begin{table}[ht]
\begin{center}

\begin{tabular}{|c|c|c|c||c|c|c|c|}
\hline
 \textbf{Configs.} & \textbf{Reliability Measures} & \textbf{By
   Simulation(hr)\footnotemark[3]} & \textbf{Time} & \textbf{By Decomp.(hr)} & \textbf{Err(\%)}  & \textbf{States} & \textbf{Time} \\
\hline
\multirow{1}{*}{single-pathing} 	&  MTTDIL	 & 5940 &  2.7 hr & 5972    & 0.54 & 13     & 137s\\ \hline
\multirow{1}{*}{multi-pathing}	&  MTTDIL 	  & 6968 &  3.6 hr & 7015    & 0.7 & 148    & 137s  \\  \hline
                                                                        
\end{tabular}
\end{center}
\caption{Results with Simulation and Hierarchical Decomposition for the systems of Fig \ref{fig5}); 
 Err: \% diff betw the two results, SMTTDIL: System MTTDIL}
\label{table51}
\end{table} 
\subsection{Hierarchical decomposition}
\label{sec4.3}
From Figs.\ref{fig5:a} and \ref{fig5:b} we note that the disks and the 
interconnects that connect the disks of a RAID group to the 
expanders contribute to the reliability of that particular RAID
group. Hence,  we can model them separately,  i.e. 
we can divide the whole system into subsystems that can be modelled
independently and the submodel results can be used in the higher level model.
Each subsystem consists of the disks and the interconnects from
expanders to the disks.  
The subsystems are logically separable rather than physically as each
of them is connected to the components shared by all of them (such as
a controller).  
This technique is called ``hierarchical decomposition'' which is a very general
technique to model large systems. Trivedi et al.  \cite{Trivedi} has
applied this technique to model and analyze the reliability of large systems 
such as telecommunication systems and cloud computing systems. The
step of dividing the system into subsystems is called the
``decomposition phase'' and the step of using the sub-model results
into the final model is called the ``aggregation phase''. 

However, our storage systems are designed with high availability in
mind with a multiplicative set of paths and are somewhat different from 
the systems considered by Trivedi\cite{trivedi-cloud} where strong modularity is present
with only a few inter-module paths.
Each of the modules in the latter
can be modelled separately and the outputs of the sub-models can be fixpoint-iterated to get the final result whereas the combinatorial structure 
of our systems (where each component designed with redundancy in mind has to
be connected to all other adjacent non-similar components for
high availability) makes it difficult to directly use Trivedi's
technique. Using tool support (such as with PRISM system) is critical
to sample the many failure paths to estimate the reliability of such
HA systems with good accuracy.


To model a system with multiple RAID groups, we model each of the RAID
groups (where a RAID group comprises the disks in it and the
interconnects that connect to the disks from expanders) separately and feed the
results to the model at a higher level. Hence, we 
have two levels: RAID group level and system level. 
If the RAID group itself is too large to model in PRISM, we use hierarchical decomposition to model the RAID group itself,  i.e. we 
model each disk (where a disk also includes its interconnects) separately and feed the results
to the model at the RAID group level. Hence,  in this case,  we have
three levels: the disk level,  the RAID group level and the system
level.

\begin{figure}
\centering
\includegraphics[height=1.7in]{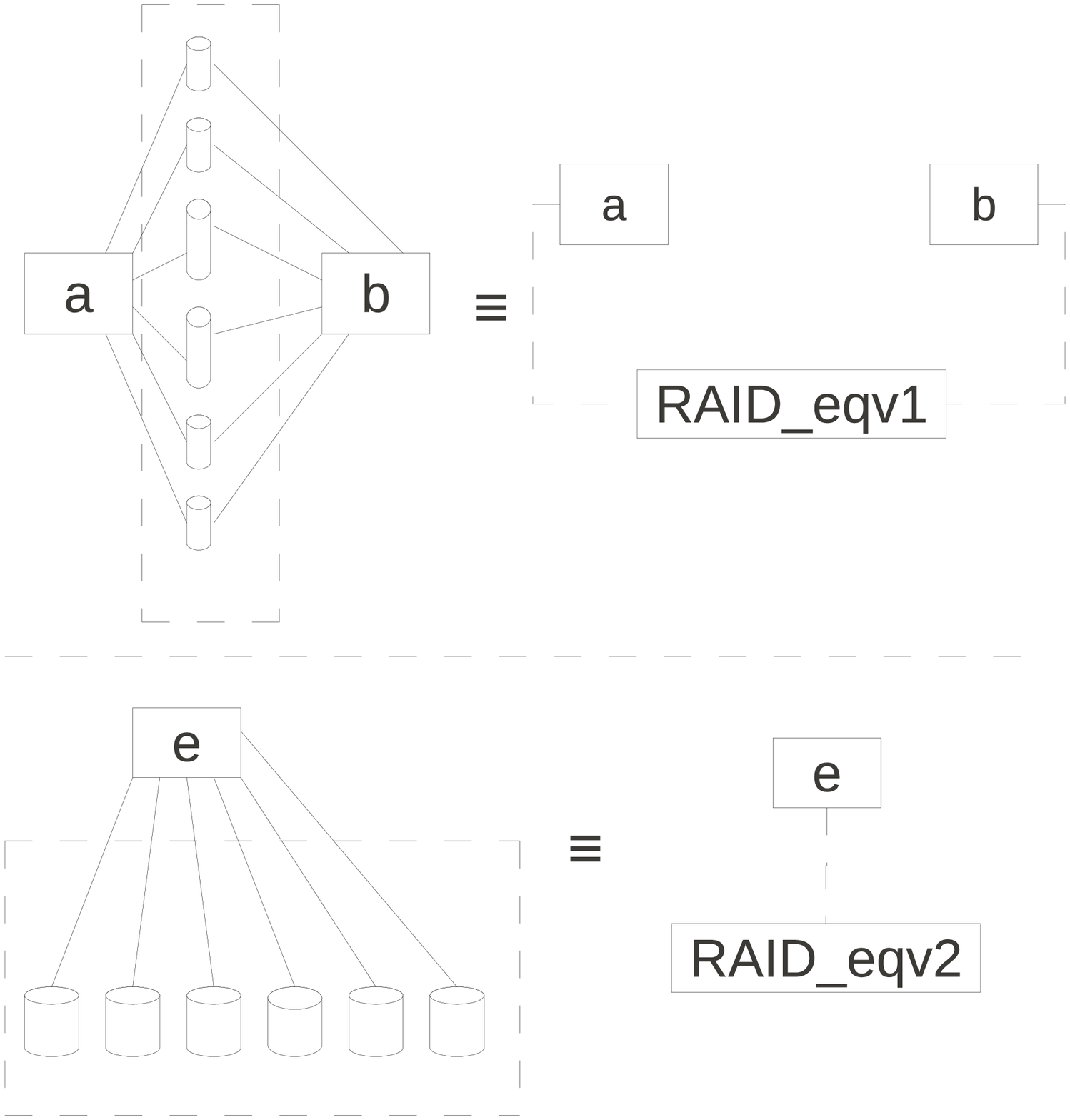}
\caption{Decomposition phase for the systems of Fig.\ref{fig5}}
\label{fig6}
\end{figure}

Note that we are using an 
approximation technique: when we reduce an $n$-state,  
$m$-transition model (representing a subsystem) to a 2-state,  1-transition model (representing the equivalent component),  some errors are
introduced.  
For example,  the holding time in the up/down state in the 2-state,  1-transition model 
is exponentially distributed,  while in the original $n$-state $m$-transition model,  that may not be true. 
The necessary conditions for this technique to be exact are:
\begin{enumerate}
 \item The computed measures are steady-state. 
 \item The subsystems transformed into ``equivalent'' model are stochastically independent 
from other subsystems. If not,  the technique can still be used,  provided that the dependence
can be expressed at some higher level in the hierarchy. 
\end{enumerate} 

As our reliability measure is ``mean'' (as opposed to steady state probability measures calculated
by Trivedi et al. for his systems),  when we use this technique 
we obtain approximate results. Moreover,  
when we separately model a subsystem,  ignoring the
complete system,  we lose some failure events. If
we do a proper decomposition,  these are rare and the accuracy is not affected.
Our results show that the accuracy 
depends on how close the subsystems (independently considered)
approach an exponential failure distribution (constant failure rate).
\begin{figure}[htp]
  \centering
  \subfigure[single-pathing]{\includegraphics[width=.49\linewidth]{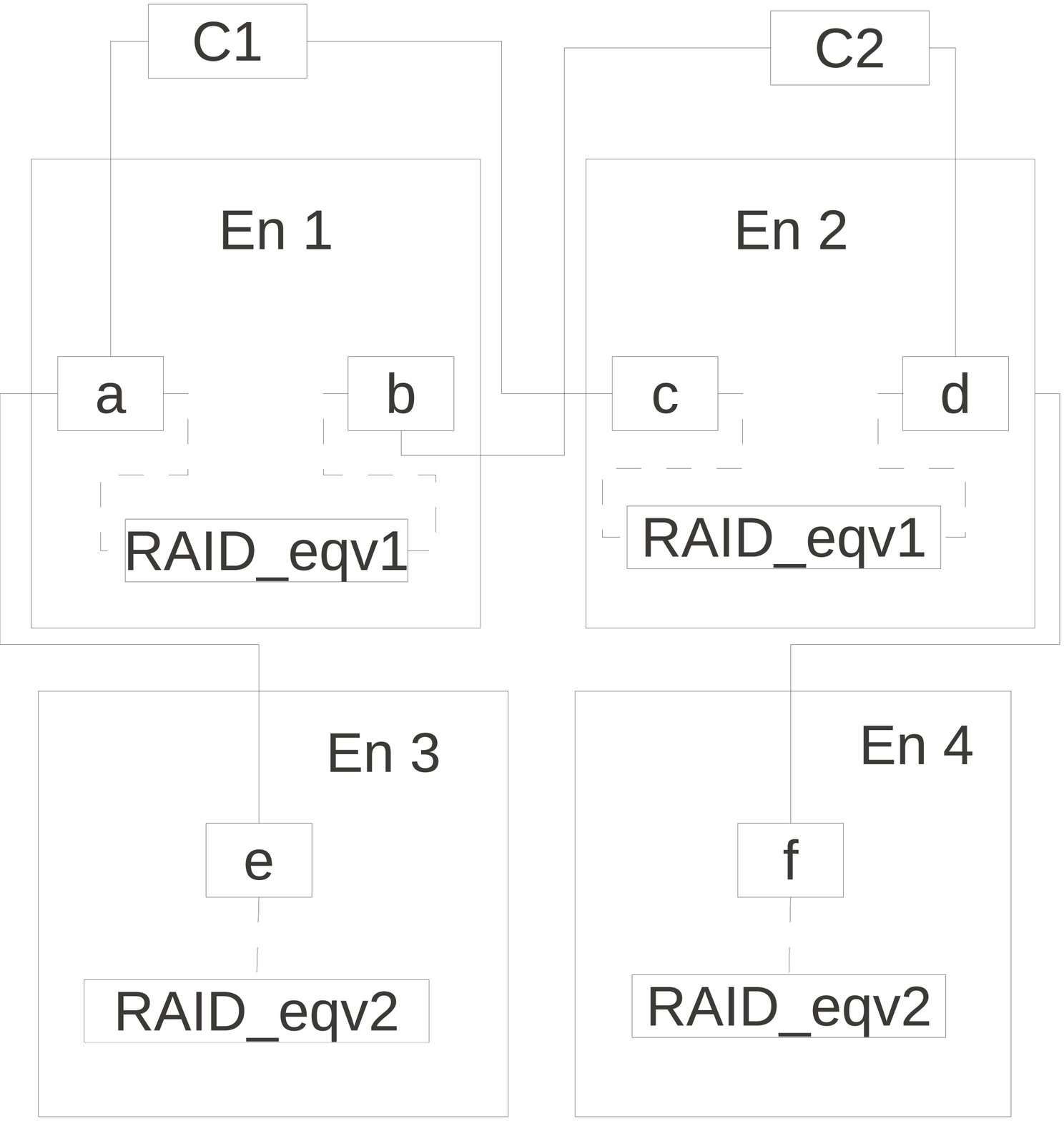}}\hfill
  \subfigure[multi-pathing]{\includegraphics[width=.49\linewidth]{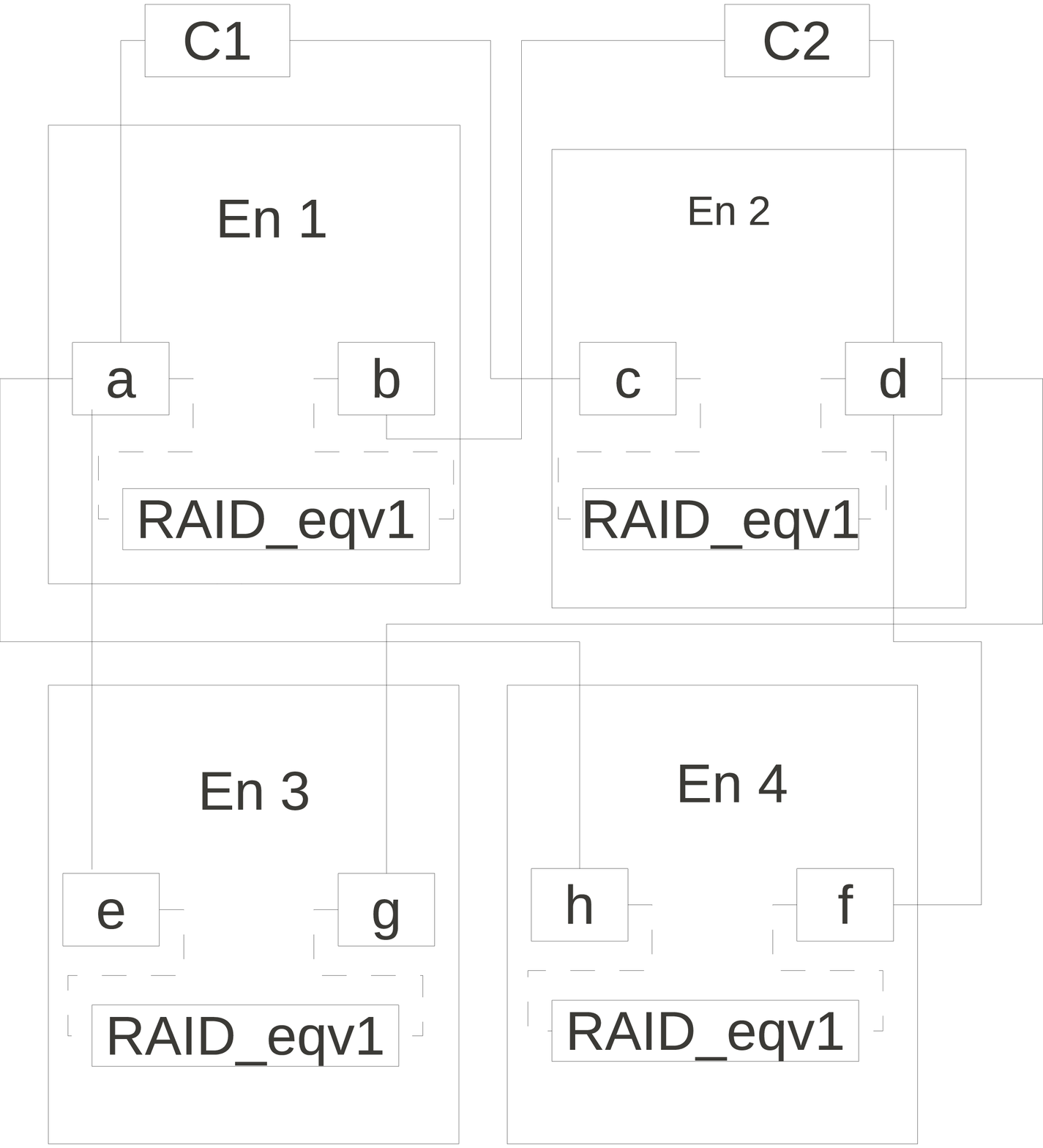}}
\caption{Aggregation phase for the systems of Fig.\ref{fig5}}
\label{fig7}
\end{figure}

\footnotetext[3]{simulation widths are almost 1\% of the point estimator}
Fig.\ref{fig5:a} shows 4 subsystems (independently considered) in the system: Sub1,  Sub2,  Sub3,  Sub4. 
Hence,  we have two level hierarchy for the systems of Fig.\ref{fig5}. In level 0 (lower level),  
we model the subsystems (independently considered) and in level
1 (higher level) we model the shared components.
Figs.\ref{fig6} and \ref{fig7} show the decomposition phase and the aggregation phase respectively. 
We calculate the reliability measures using hierarchical decomposition and
the results we obtain is consistent with simulation results (Table \ref{table5}). 
We use a shell script to calculate reliability measures for a system using hierarchical decomposition. In the script, 
lower level model results are calculated first using PRISM
and then passed as a input parameter to the next higher level model and so on.

Table \ref{table51} shows the results of hierarchical
decomposition. 
Our results show that the use of hierarchical decomposition is indicated in the following three scenarios:
\begin{enumerate}
 \item The system is too large to simulate and there is no other option except using hierarchical decomposition. 
 \item When the system can be simulated but each subsystem
   considered independently has a constant failure rate (which can be
   checked using goodness-of-fit tests such as Kolmogorov-Smirnov
   test).  
 \item When each subsystem has a small contribution to the
   reliability of the overall system (which can be checked by
sensitivity analysis).  
\end{enumerate}
\subsection{Modelling of some known field configurations}
\label{sec4.4}
Tables \ref{table6} and \ref{table7} shows the MTTF of the components used in 
some large storage configurations and field MTTDIL value\footnote[4]{The
details about the field data are not known (for example, how many samples are used to get the mean values of Table \ref{table7}
for these configurations).} respectively. 
In these systems, a RAID5 group consists of 6 disks.
For the systems with multiple controller pairs,  no RAID group is implemented across controller pairs.
We model these systems and check whether the model results match the field data.
\begin{table}[ht]
\begin{center}
\begin{tabular}{|c|c|}
\hline
 \textbf{Components} & \textbf{MTTF (in hr)} \\
\hline
Controller 	                        &	      35000\\ \hline
Type1 enclosure(t1)                         &          60000 if $\leqslant$ half-full; else 23000 \\ \hline
Type2 enclosure(t2)                         &             50000 when full\\ \hline
\end{tabular}
\end{center}
\caption{MTTF of the components for large storage configs.}
\label{table6}
\end{table}
\subsubsection{Modelling single controller-pair systems}
\label{sec4.4.1}
\begin{enumerate}
\item We simulate the 24 disk system of Table \ref{table7} in PRISM simulator 
with 99\% C.I. and $10^{6}$ samples. The time for simulation is 
5 hr with MTTDIL = 45859 hr.
The subsystem consisting of 24 disks and 48 
interconnects contributes very little to the MTTDIL of the whole
system; this has been verified using sensitivity analysis.
Using hierarchical decomposition, we get MTTDIL as 46098 hr with time
for model checking being only 4 min. 
\item 
We use hierarchical decomposition to model the 60 disk system of Table \ref{table7} because it is too large to 
simulate in PRISM. MTTDIL is 20960 hr with time taken for model
checking also being 4 min.    
\end{enumerate}
\subsubsection{Modelling multiple controller pair systems}
\label{sec4.4.2}
Fig.\ref{fig8:a} shows a configuration with multiple controller pairs. We can think of these systems consisting of 
subsystems each corresponding to one controller pair and not connected to each other. Hence,  they are ``totally independent'' or ``physically independent'' (they do not share any common components). 
Here these subsystems are also symmetric.
There are various approaches to model such systems:
\begin{figure}[htp]
  \centering
  \includegraphics[width=2in]{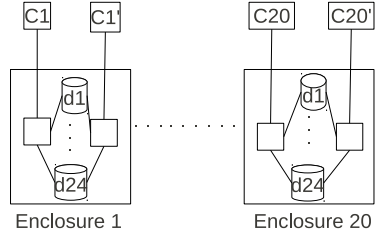}
\caption{20 controller pair 20 enclosure 480 disk systems}
\label{fig8:a}
\end{figure}
 
\textbf{Hierarchical Decomposition}: We model/simulate any one ``totally independent subsystem'' and 
do a hypothesis test to check whether a subsystem has a constant failure 
rate or not. If yes,  then MTTDIL of the whole system 
is = MTTDIL of a subsystem / number of ``totally independent'' subsystems. 

If we can model a ``totally independent subsystem'' using PRISM model
checker or discrete-event simulator, we call this 
``partial hierarchical decomposition'' (P).  If a ``totally
independent subsystem'' is too large to model using PRISM simulator we
use hierarchical decomposition to model the subsystem itself; we 
call it ``total hierarchical decomposition'' (T). We have used both
these techniques but due to lack of space we do not present results here.


\textbf{Simulation of the whole system}: When each of the independent subsystems 
has a non-constant failure rate,  then we can simulate the whole system
to get a confidence 
interval for our reliability measure.
Let $Y$ be the random variable to
denote the time to DIL (data inaccessible or data loss)
of the whole system and $X_{1},  X_{2},  \dots, X_{n}$ be the time to DIL
of each of the $n$ independent subsystems. Then $Y=min(X_{1}, X_{2},  \dots, X_{n})$. 
To calculate $E[Y]$ we take $N$ observations. In the
$i$-th observation we simulate each of the subsystems and get the samples 
$x_{i1},  x_{i2}, \dots, x_{in}$ ($x_{ij}$ is the time to DIL of the $j$-th subsystem in $i$-th observation) and calculate $y_{i}=min(x_{i1}, x_{i2}, \dots, 
x_{in})$. Estimated $E[Y]$ = $\sum_{i=1}^{N} y_{i}/N$ 

\begin{figure}[htp]
  \centering
  \subfigure[Upper Sum]{\includegraphics[width=.49\linewidth]{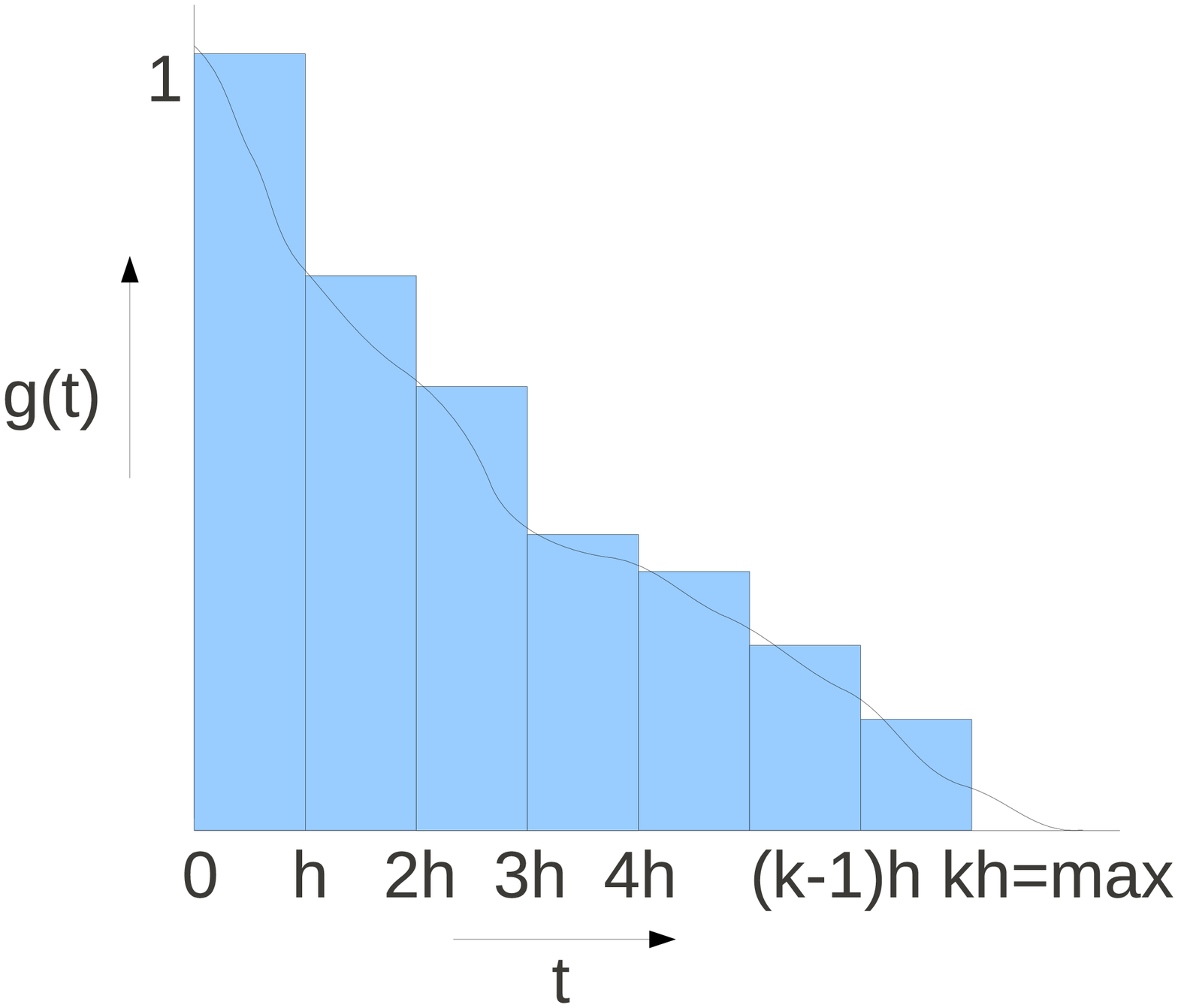}}\hfill
  \subfigure[Lower Sum]{\includegraphics[width=.49\linewidth]{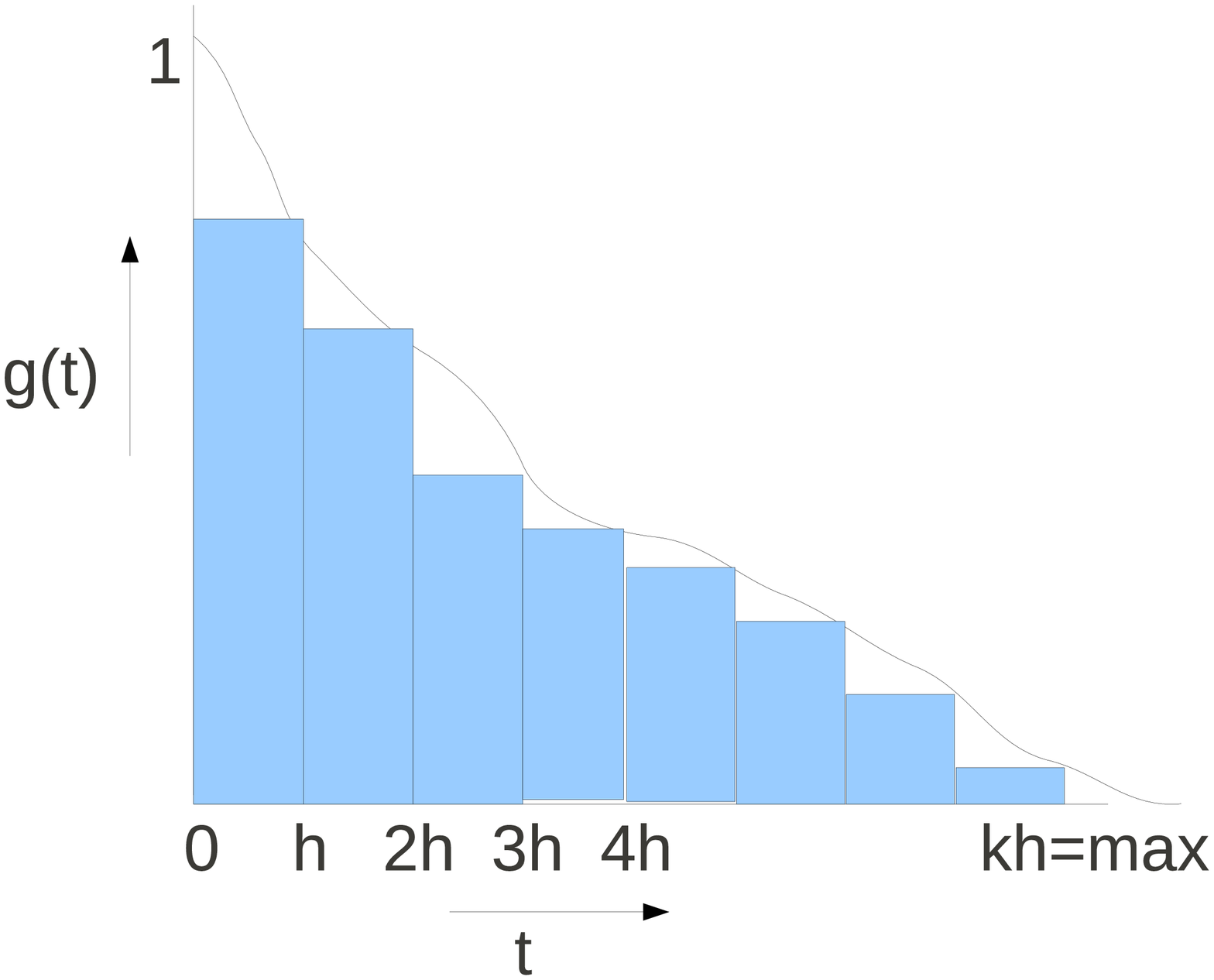}}
\caption{Discretization approach}
\label{fig9}
\end{figure}

\textbf{Discretization process}: The ideal approach to
model such large systems is to
calculate the probability of DIL for a ``totally independent subsystem'' 
and calculate the probability of data inaccessibility or loss 
for the whole system using the formula 
\begin{align}
W(t)= 1-(1-F(t))^{n}.
\end{align}
where  
$W(t)$ is the probability of DIL of the whole system, $F(t)$
is the probability of DIL of a subsystem and 
$n$ is 
the number of independent subsystems present in the system. Now, 
\begin{align} 
E[X]=\displaystyle\int_{0}^{\infty}(1-W(t))dt=\displaystyle\int_{0}^{\infty}(1-F(t))^{n}dt. 
\end{align}
Let $g(t)=(1-F(t))^{n}$.
We calculate MTTDIL from this probability values using sampling based 
techniques as follows:
First,  we find a maximum range of the random variable $X$ which represents time to DIL. 
For that we find the value $max$, such that $g(max) < \delta$ where 
$\delta$ is some error bound,  say,  1E-4. Then we divide the range of the random variable ($0$ to $max$) into steps each of size $h$. 
\begin{align}
U &= hg(0)+hg(h)+\dots+hg((k-1)h) \label{eqn1}\\
L &= hg(h)+hg(2h)+\dots+hg(kh) \label{eqn2}
\end{align}
where $U$ and $L$ stands for Upper and Lower Riemann sum respectively. $k$ is the number of steps each of size $h$,  so $kh=max$. Clearly, 
\begin{align}
 L < E[X] < U
\label{eqn3}
\end{align}
To get a good bound on $E[X]$,  our aim is to reduce the difference between $U$ and $L$. Suppose,  we want a difference of $\epsilon$  
i.e. $U$-$L$ = $\epsilon$. Then
\begin{align}
 U-L=h[g(0)-g(kh)]=h[1-g(kh)] \approx h 
\end{align}
assuming $g(kh)=g(max)\approx0$.
Hence,  we 
choose a step size ($h$) of $\epsilon$.
The process is shown in Fig.\ref{fig9}.

For both simulation and discretization approach, we use a shell script 
to calculate the MTTDIL values. For simulation, the 
script generates samples (using PRISM simulator) for time to DIL 
for each ``totally independent subsystem'' and takes minimum of them, repeats 
the whole process for the specified number of observations and 
calculates mean and variance of those minimum values from 
each observation. In case of discretization approach, the 
script generates probabilities of DIL for a ``totally independent 
subsystem'' at time $t$ where $t$ ranges from 0 to $max$ with a 
step size of $h$ using PRISM simulator. From these probabilities, 
we calculate MTTDIL with the sampling technique described above 
using a C program invoked from the script. 
 
Table \ref{table8} under the column labeled ``without correlated failure'' shows the results 
after applying these approaches to the systems of Fig.\ref{fig8:a}. 
We assume $\epsilon$ = 175 hr for the discretization approach.
\paragraph{Comparison between the techniques:}
The results using hierarchical decomposition are slightly off from the results using other two approaches. 
The reason is the constant
failure rate assumption of each individual subsystem. We are able to reject the 
hypothesis in each of the above cases that each individual
subsystem has a constant failure rate using Kolmogorov-Smirnov test with 5\% level of significance. 
If each of the individual subsystem has a failure distribution which is far from exponential then it is 
best to use the simulation approach. The method of calculating mean using discretization approach
has an advantage in that the error is bounded (with probability 1) but it has the following
disadvantages:
\begin{enumerate}
 \item Each individual subsystem is modelled using PRISM
   simulator. Hence,  the $F(t)$ result value has some error. When
we calculate $W(t)=1-(1-F(t))^{n}$, the error can increase. 
\item To get a small difference between $U$ and $L$, we need a small value for step size $h$ i.e. large number of steps. Hence,
this procedure can take a long time for small $h$. 
\end{enumerate}
\subsubsection{Comparison of model results with field data:}
\label{sec4.4.3}
Our computed results, however, deviate significantly from the field data. The possible reasons are:
\begin{enumerate}
 \item Since we do not have field data for disk failure, we may have assumed a simple model for disk 
       failure instead of, for example, Weibull. Disk failure model
       may affect the result because the number of disks present in the system is 
       much higher than other components.
 \item Correlated/burst failure of disks: Many disks in an enclosure may fail within a short span due to high 
temperature,  power supply spikes,  vibration etc. thus 
causing double disk failures almost simultaneously. 
\end{enumerate}
To identify the main factors, 
we consider other disk failure models (such as Weibull disk model or
correlated disk failure model) 
to check whether these models agree with field data. 
\begin{table}[ht]
\begin{center}

\begin{tabular}{|c|c|c||c|c||c|c||c|c|}
\cline{2-9}
\multicolumn{1}{c|}{} & \multicolumn{4}{c||}{\textbf{without correlated failure}} & \multicolumn{4}{c|}{\textbf{with correlated failure}}\\
\cline{2-9}
\multicolumn{1}{c|}{} & \multicolumn{2}{c||}{\textbf{480 disk}} & \multicolumn{2}{c||}{\textbf{600 disk}} & \multicolumn{2}{c||}{\textbf{480 disk}} & \multicolumn{2}{c|}{\textbf{600 disk}}\\
\hline
\textbf{Methods} & \textbf{MTTDIL (hr)} & \textbf{T} & \textbf{MTTDIL (hr)} & \textbf{T} & \textbf{MTTDIL (hr)} & \textbf{T} & \textbf{MTTDIL (hr)} & \textbf{T}\\
\hline
M1                                    &	2293        &	5h &  	2590       &	1.11h &	1750        &	25h   &	1253      &	3h\\ \hline
M2                                      &       2304       &    4m &      2616       &    4m &       1800       &    10m &       1290      &    10m \\ \hline
M3                                                              &       2160$\pm$109 &  29h &  2337$\pm$114 &  36h   &       1600$\pm$84 &   38h  &       1128$\pm$ 48 & 29h \\ \hline
M4                                                 &       $L$=2057, $U$=2232 & 48h & $L$=2308, $U$=2483 & 48h &  $L$=1510, $U$=1685 & 36h  & $L$=1062, $U$=1237 & 31h \\ \hline
\end{tabular}
\end{center}
\caption{Model results for multiple controller pair configs.: M1:
Hierarch. Decomp. (P), M2: Hierarch. Decomp. (T), M3: Simulation, 
M4: Discretization; T: Time. The field data for 480 disk configs. is
\textbf{1700 hr} and for 600 disk configs. is \textbf{1200 hr}.}
\label{table8}
\end{table}
\section{Detailed model of disk subsystems}
\label{sec5}
\subsection{Disk reliability model with Weibull distribution}
\label{sec5.1} 
We use the detailed disk reliability 
model of Elerath et al. \cite{disk3}. The model assumptions are as follows:
Time to operational failure (TTOp) with a 2-parameter Weibull (shape = 1.12,  scale = 461386 hrs); 
Time to restore (TTR) with a 3-parameter Weibull (shape  = 2,  scale = 12 hours and offset 6 hours); 
Time to scrub (TTScr) with a 3-parameter Weibull (shape  = 3,  scale = 168 hours and offset 6 hours); 
Time to latent defect (TTLd) with shape = 1 (an exponential distribution) and scale = 9259 hours

Elerath et al. presented a sequential Monte Carlo simulation method using these models 
to calculate DDF($t$) where DDF($t$) stands for number
of double disk failures in time $t$. Elerath also created a simple DDF(t) equation \cite{ddf_eqn} for N+1 RAID
systems to calculate expected number of double disk failures. A DDF occurs when any two disks of a RAID5 group experience operational failure or one disk has a latent defect followed 
by operational failure from another disk. As PRISM does not 
support anything other than exponential distributions, we approximate Weibull distributions using phase type distributions (sum of exponentials). 
All of the above Weibull failure/repair models have increasing failure rates.
 We use the same 3 state model of \cite{infant_mortality} for each 
of the Weibull models and find the parameters of the models using the standard 
technique of moment matching. The pdf (probability density function) of the fail state in the 3-state
model is:  
$$\dfrac{1}{\sigma+\alpha-\beta} [\beta \sigma e^{-\beta t} + (\alpha-\beta)(\sigma+\alpha)e^{-(\sigma+\alpha)t}]$$
The first three moments of this distribution are:
\begin{align}
\mu_1&=\frac{\frac{\sigma}{\beta}+\frac{\alpha-\beta}{\alpha+\sigma}}{\alpha-\beta+\sigma} ;~~~~ 
\mu_2=\frac{\beta^2+\sigma(2 \alpha+\sigma)}{\beta^2 (\alpha+\sigma)^2} ;~~~~
\mu_3=\frac{2 \left(1+\frac{-\alpha^3+\beta^3}{(\alpha+\sigma)^3}\right)}{\beta^3}.
\end{align}

Solving these three equations,  we obtain $\sigma$, $\alpha$ and $\beta$: 

$\sigma = \frac{4 \mu_1^3-6 \mu_1 \mu_2+\mu_3 \pm \sqrt{x}}{y} ;~~ 
\alpha = -\frac{2 \left(\mu _1^3-3 \mu _1 \mu _2+\mu _3\right)}{y} ; ~~
\beta =  \frac{2 \mu _1^3- \mu _3 \mp \sqrt{x}}{y} \ where\\
x=\scriptstyle {-2\mu_1^6+6\mu_1^4\mu_2-18\mu_1^2\mu_2^2+18\mu_2^3+8\mu_1^3\mu_3-12\mu_1\mu_2\mu_3+\mu_3^2} ;~~
y=\scriptstyle {\mu_1^4+3\mu_2^2-2\mu_1\mu_3}\\$

We equate them with the first three moments of Weibull for each of the three cases: TTOp,  TTScr,  TTR. For TTOp, 
the solutions turn out to be 
$\alpha = 1.72E-6$ and either 
 $\sigma = 2.49E-6$,  $\beta = 2.88E-6$ or,  equivalently
 $\sigma = 1.16E-6$,  $\beta = 4.21E-6$
\subsubsection{Comparison of approx. model with Weibull}
\label{sec 5.1.1}
\begin{figure}[htp]
  \centering
  \subfigure[Pdf functions]{\label{fig320 :a}\includegraphics[width=0.49\linewidth, height=2.5 in]{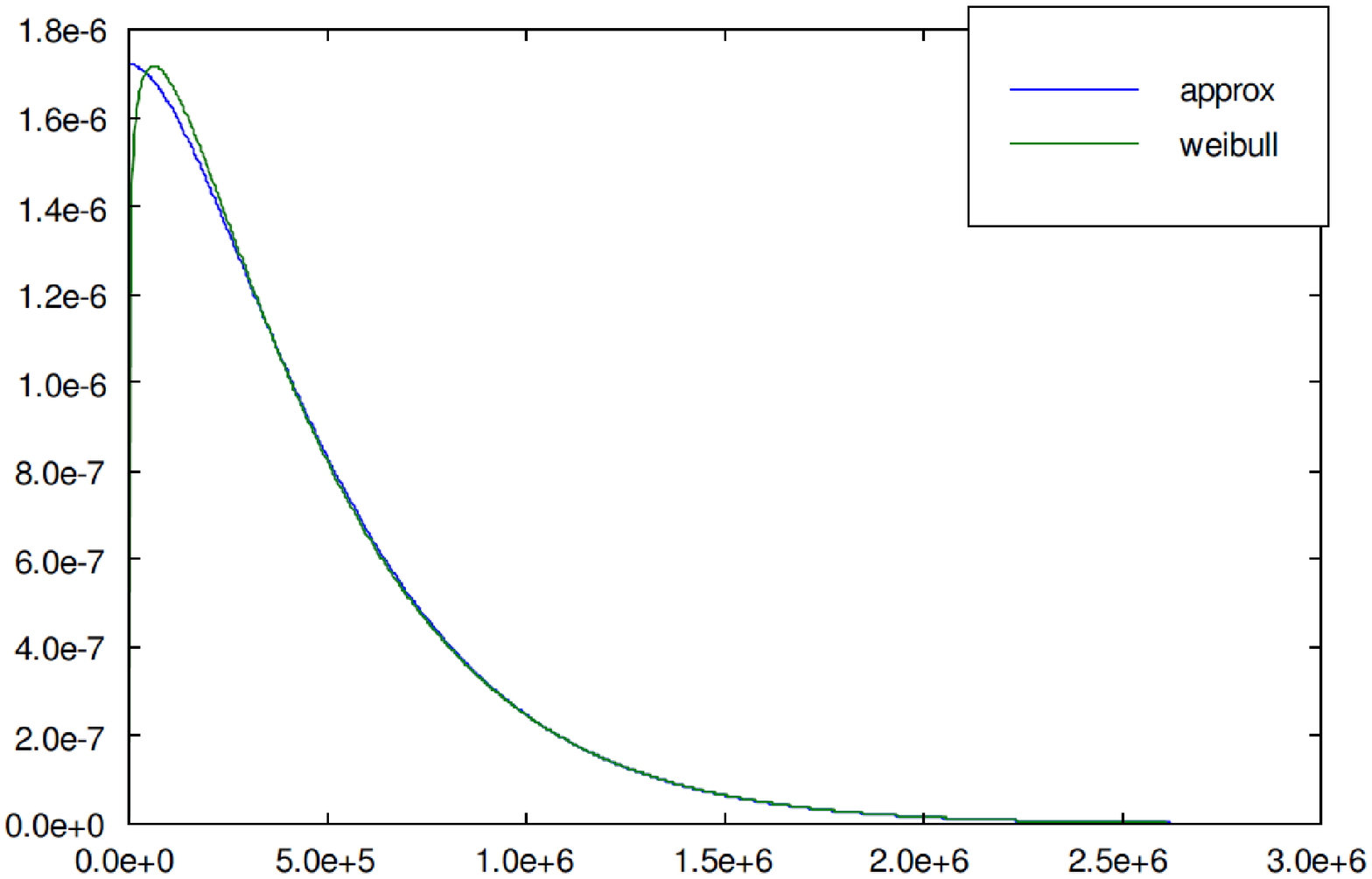}}
  \subfigure[Hazard functions]{\label{fig320 :b}\includegraphics[width=0.49\linewidth, height=2.5 in]{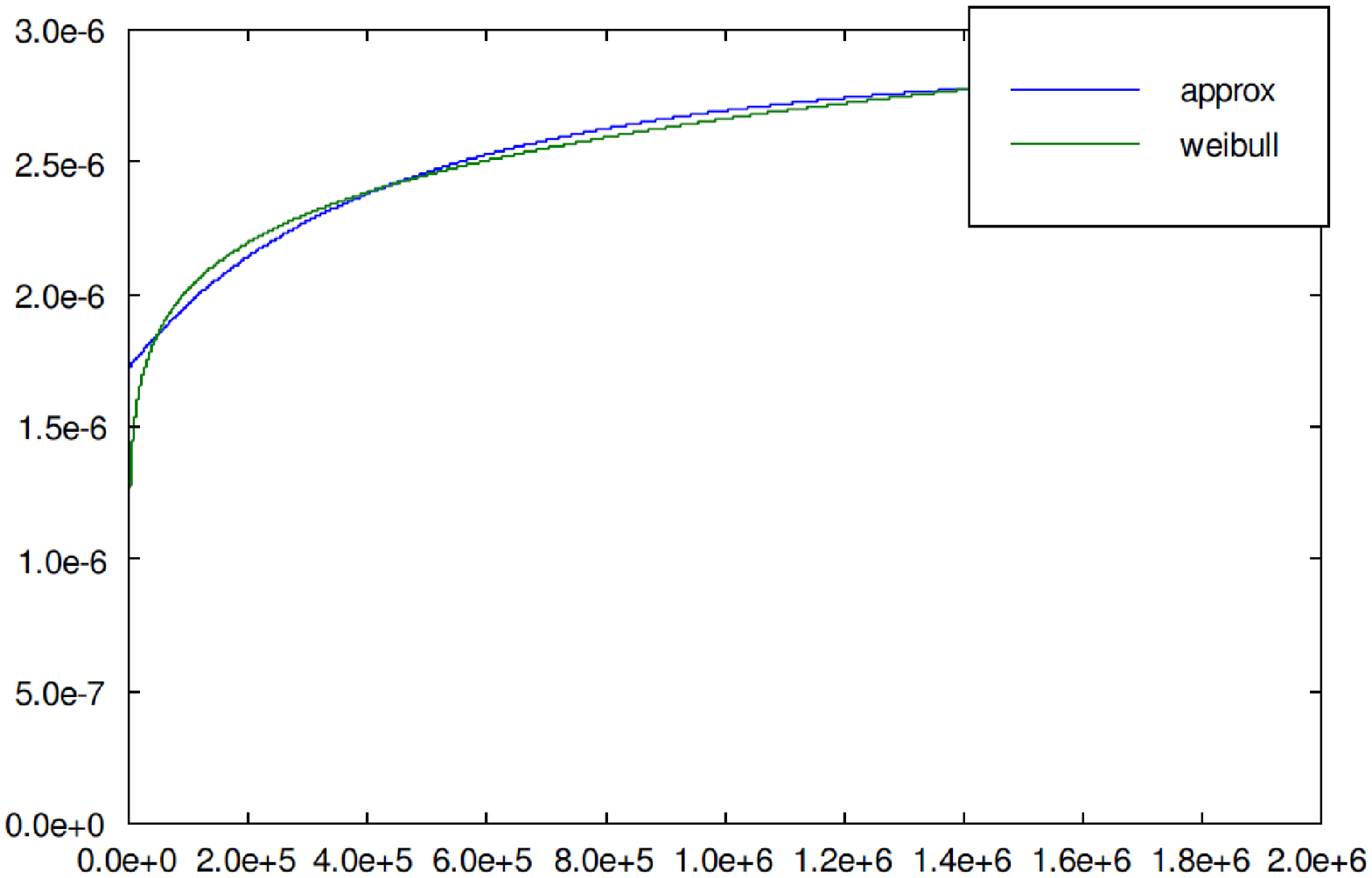}}
\caption{Approximate vs. Weibull; X axis shows time in hrs}
\label{fig320}
\end{figure}

To check how well this pdf approximates Weibull distribution, we compare the pdf and hazard functions of approximate and Weibull models 
(Figures \ref{fig320}). The hazard rate for the approximate model becomes constant 
after some time. This can be understood by looking into the slope of the hazard rate function for the approximate model :
$$\frac{\sigma(\beta-\alpha)e^{-(\sigma+\alpha+\beta)t}}{(\frac{\sigma}{\sigma+\alpha-\beta}e^{-\beta t}+
\frac{\alpha-\beta}{\sigma+\alpha-\beta}e^{-(\alpha+\sigma)t})^2}$$
Note that the slope function is a non-negative decreasing function for $\beta > \alpha$. Hence after some time slope becomes zero. 
 
To understand the differences better, we look at the differences between the two CDFs (Approximate minus Weibull). 
The difference is never more than +0.006 or less than -0.003.  Therefore, when using the CDFs to compute probabilities of any interval,
 the results will never be erroneous by more than 0.006 - (-0.003)  = 0.009, less than 1\%. The differences in the right tails apparently 
become zero, indicating the approximation to be very good for right tail probabilities.
\begin{figure}[htp]
  \centering
  \includegraphics[width=2.3 in]{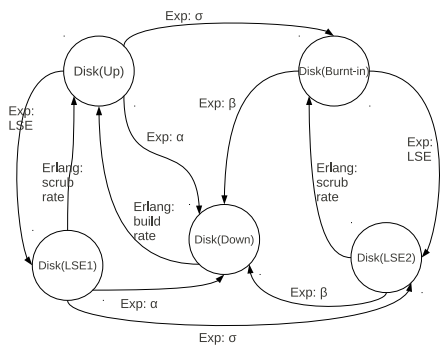}
\caption{Approx. Disk model based on Gopinath et al.  \cite{disk2}: one
  difference is that we consider here a more accurate model that has a transition from Disk(LSE1) state to the 
Disk(LSE2) state with rate $\sigma$ rather than a transition from Disk(LSE1) to Disk(Burnt-in) state.}
\label{fig32 :b}
\end{figure}
 
For TTOP and TTScr, with the same approach, we get complex number for $\sigma$ and 
$\beta$ and negative value for $\alpha$ for each of the 
two solutions respectively. Hence,  we use other phase type distributions 
such as Erlang distributions  \cite{disk2}. We use a 3-stage Erlang model. For TTScr
$\lambda$ = 0.019228232 and for TTR $\lambda$ = 0.180345653. 
Using these models for each type of failure/repair we build a detailed disk model (Fig.\ref{fig32 :b}). 

\paragraph{Comparison of PRISM, Monte Carlo Simulation and DDF(t) equation Results:}
We compare the reliability of RAID subsystems using PRISM model, Monte Carlo Simulation 
and DDF(t) equation (Table \ref{table18} and Table \ref{table20}). 
We try to keep the variance of both PRISM and Monte Carlo Simulation 
results same so that we can make a fair comparison. Hence, we 
set the termination epsilon parameter in case of PRISM and the number of 
experiments parameter in case of Monte-Carlo simulation  accordingly.  
Results from Table \ref{table18} and \ref{table20} (under the column with 3-state disk failure model) show that 
DDF(t)  values calculated from PRISM model are similar with that of the MC-Sim and DDF(t)  equation. Due to the 
front-overloading of our approximate pdf (compared to the actual Weibull pdf), the difference between DDF(t)  values calculated 
using PRISM and the other two methods (MC-Sim and DDF(t) equation) is much higher in the beginning. 
\begin{table}[ht]
\begin{center}
\begin{tabular}{|c|c|c|c|c|c|c|c|c|}
\hline
 \textbf{Time(yr) } & \textbf{pDDF$_{3}$(t)} & \textbf{pDDF$_{4}$(t)} & \textbf{sDDF(t)} & \textbf{eqDDF(t)} & \textbf{sDev$_{3}$(\%)} & \textbf{sDev$_{4}$(\%)} & \textbf{eDev$_{3}$(\%)}   & \textbf{eDev$_{4}$(\%)}\\
\hline
 1                                   &	7.12    & 5.59   &  5.63   &   5.64 & 26.5 & -0.72 & 26.24 & -0.9\\ \hline
 2                                   & 14.37    & 12.2   & 12.23   &   12.26 & 17.5& -0.21  & 17.21 & -0.46\\ \hline
 3                                   & 21.67    & 19.26   & 19.21 &    19.31 & 12.8& 0.28  & 12.22 & -0.24\\ \hline
 4                                   & 28.99    & 26.59   & 26.43 &    26.64 &9.7  & 0.59  &  8.82 & -0.20  \\ \hline
 5                                   & 36.35    & 34.06   & 33.8 &     34.21 &7.5  & 0.75 &   6.26 & -0.45 \\ \hline
 6                                   & 43.73    & 41.6   & 41.27 &     41.96 & 6   & 0.8   &   4.22 & -0.86 \\ \hline
 7                                   & 51.13    & 49.17   & 48.79  &   49.87 & 4.8 & 0.77 &  2.53 & -1.41\\ \hline
 8                                   & 58.54    & 56.73  &  56.36 &     57.91 & 3.9 & 0.66 &  1.09 & -2.09 \\ \hline
 9                                   & 65.96    & 64.27  &  63.93 &     66.08 &3.2 & 0.57 &   -0.18 & -2.73\\ \hline 
10                                   & 73.39    & 71.78  &  71.50 &    74.35 & 2.7 & 0.38 &  -1.29 & -3.46 \\ \hline
\end{tabular}
\end{center}
\caption{DDF(t)  per 1000 RAID groups for 6 disk RAID5  : 
PRISM Model (PRISM DDF(t))  vs. Simulation (sDDF(t))  vs. DDF(t)  equation (eqDDF(t)) result;
pDDF$_{i}$(t)= DDF calculated in PRISM using $i$-state disk failure model.
sDev = Deviation of PRISM results from Simulation results;
eDev = Deviation of PRISM results from DDF(t)  equation results;
Time taken for Model Checking = \textbf{37 sec} (using 3-state model) and \textbf{4.3 min} (using 4-state model) 
 while time for Simulation = \textbf{8 min}; both PRISM and simulation error are 1\%;
}
\label{table18}
\end{table}
\begin{table}[ht]
\begin{center}

\begin{tabular}{|c|c|c|c|}
\hline
 \textbf{Time(yr) } & \textbf{PRISM DDF(t) } & \textbf{sDDF(t) } & \textbf{sDev(\%) }\\
\hline
 1                                   &	2.26       &  1.92  &  17.7\\ \hline
 2                                   &  4.62       &  3.84   & 20.3  \\ \hline
 3                                   &  7.03       &  6.46 &  8.8 \\ \hline
 4                                   &  9.51       &  9.32 &  2    \\ \hline
 5                                   & 12.04       & 12.16 & -1\\ \hline
 6                                   & 14.63       & 14.87 & -1.6 \\ \hline
 7                                   & 17.27       & 18.24  & -5\\ \hline
 8                                   & 19.96      &  21.52 & -7.3 \\ \hline
 9                                   & 22.71      &  24.56 & -7.5 \\ \hline 
10                                   & 25.50      &  28.16 & -9.4 \\ \hline

\end{tabular}
\end{center}
\caption{DDF(t)  per 1000000 RAID groups for 8 disk RAID6
    with 3-state model: 
Time taken for Model Checking = \textbf{12.6 min} while time for Simulation = \textbf{26 hr}; 
PRISM error is 1\% and Simulation Error is 4\%}
\label{table20}
\end{table}

It can be noted that 
the higher deviation between the results of PRISM and simulation due to front overloading of the 
approximate pdf can be reduced by adding more states in the Markov model. We consider a 4-state model 
to check how well it approximates Weibull. 
Note that a 4-state Markov model has 5 model parameters. To estimate them using 
moment matching will be a very hard problem. Hence we try to estimate the parameters by trial and error method.

Next, we check how this 4-state 
model performs when modelling disk subsystems. Table \ref{table18} (under the column 4-state disk model) shows the DDF(t) values 
computed using the 4-state model and how it agree with simulation and DDF(t) equation results. Note that in the time period 
of $t$ = 0 to 10 yr, the deviations are now much less (especially in the initial period), but the hazard rate function starts to flatten 
much earlier compared to the 3-state model with moment matching. 
As we calculate 
mean for the whole system, we use the 3-state model with parameters 
estimated using moment matching to approximate Weibull model 
for modelling the whole system. 

Apart from the results of Table \ref{table18} and \ref{table20}, our Weibull approximation produces results which are in reasonable agreement with Greenan's simulation results \cite{Greenan} (details can be found in
Appendix \ref{app3}).  
\subsection{Whole system modelling with detailed model}
\label{sec5.2} 
Assuming the detailed disk 
model, we can model the large storage systems (Table \ref{table7}) using hierarchical decomposition. 
Table \ref{table7} shows the results using this detailed disk model of Figure \ref{fig32 :b}. 
The model results, however, still deviate from the field values,
although they are closer to the field data than the results with simple 3
state disk model.
\begin{table}[ht]
\begin{center}
\begin{tabular}{|c|c|c|}
\hline
 \textbf{Configs.} & \textbf{Model result(hr)} & \textbf{Field value(hr)} \\
\hline
(1, 1, t2, 24)	       &  41396          &	35000     \\ \hline
(1, 1, t1, 60)         &  18563          &       11000     \\ \hline
(20, 20, t2, 480)      &  2070           &        1700     \\ \hline
(20, 20, t1, 600)      &  2253           &        1200     \\ \hline

\end{tabular}
\end{center}
\caption{Weibull Model results vs. field value for the large storage configurations. 
Configs. $(a, b, c, d)$ $a$: number of controller-pairs, $b$: number of enclosures, 
$c$: type of an enclosure,  $d$: number of disks}
\label{table7}
\end{table}

Hence,  we postulate correlated failure as a possible reason for the difference between model 
results and real system field results. Garth et al.  \cite{disk_mttf}
has found existence of strong correlation in disk failures. The study by 
Jiang et al.  \cite{netapp_fast} also show the existence of correlated
failure for disks. They show that RAID group spanned 
across multiple enclosures exhibit lower correlated failure. The 
study also shows that the amount of correlation depends on the 
particular enclosure model. We show that our correlated failure model is consistent 
with these findings. 
\subsection{Modelling correlated failure for disks}
\label{sec5.3}
The key point in modelling correlated failures for disks is that failure events are 
independent of each other rather than individual disks failing
independent of each other. Each failure event may involve multiple
disks. 
Let $p$ be the probability that when a disk fails another disk also fails 
simultaneously. The value of $p$ depends on the source of correlation
(for example, the enclosure).
   

Fig.\ref{fig12} shows the CTMC model for a RAID5 group consisting of $n$ disks 
in an enclosure.

We use the ``synchronized action'' language construct in PRISM to
build the model in PRISM: such an action can be used to force two or
more modules to make transitions simultaneously with a rate that is product 
of two rates. For our models,  we use one action as the ``active action''
that actually defines the rate for the synchronized transition and the other 
one as the passive action with rate as 1. 

We model the storage configurations (Table \ref{table7}) using the revised model and compare them with field data.

\subsubsection{Validation}
 As we do not have any information on $p$ for our systems, we assume a particular disk failure model (i.e. 3-state model 
or Weibull model),
estimate $p$ for each type of enclosure from the field data of 1 enclosure
configurations 
and then use that $p$ to check whether the model results match field value 
for the multiple controller pair configurations. 

The model results are given in Table \ref{table8} under the column
with the heading ``with correlated failure''; these agree with field
data quite well. 

Modelling the large storage 
configurations assuming Weibull model along with correlated failure for disks 
takes around 1.5 hr.

\section{Conclusions}
\label{sec6}
We have presented several approaches for reliability modelling of RAID storage systems
starting from 4 to 600 disks. Using these models we are 
able to perform sensitivity analysis, make cost-reliability trade-offs and 
choose better reliability configurations. 
To the best 
of our knowledge, there has been no comparable work in the open literature. 

\appendix
\section{Other reliability measures}
\label{app1}
Generally, a RAID5 group consists of 6-8 disks because a larger number
of disks increase the chance
of latent sector error during reorganization. Hence one RAID5 group
may not be sufficient to store large amounts of data. 
For systems with multiple RAID groups accessed by multiple users
we can define reliability metrics such as the following:
\begin{defi}
 \textbf{[$k$\%System MTTDIL+R(-R)]}
        Mean time before which $k$\% of ``all'' the RAID groups
        experience ``data inaccessibility or data loss'' even with (without) repair. With k=50,  this is system
        ``half-time'' when repair is (is not) possible from the ``data inaccessible'' state.
\end{defi}


In a multiple RAID system, the first measure (MTTDIL) is not
sufficiently informative as a sysadm may be interested in knowing how
many RAID groups are available for allocation to different users when
some RAID groups are undergoing
repair (either connectivity or rebuild). With the new measures, a sysadm will be able to select repair rates
and disk replacement rates to ensure satisfactory allocations of RAID
groups (for example, in a ``cloud'' setting). 
Consider also two users A and B where the disks for A are in very highly
unreliable enclosures while those for B are not.
Although the MTTDIL of the system is very low,  B experiences good reliability (as the RAID groups used by B experience high MTTDIL). 
The $k$\%System MTTDIL-R metrics can reflect this information because 
it consider failures from ``data
inaccessible or data loss'' state and hence not much affected by a single weak link in the system. Therefore it can represent the reliability 
of a system much better. 

For example, for the systems of Figure \ref{fig5} we claculate the 50\% and 100\% reliability metrics as shown in the following 
table.  

\begin{table}[ht]
\begin{center}

\begin{tabular}{|c|c|c|c||c|c|c|c|}
\hline
 \textbf{Configs.} & \textbf{Reliability Measures} & \textbf{By
   Simulation(hr)\footnotemark[3]} & \textbf{Time} & \textbf{By Decomp.(hr)} & \textbf{Err(\%)}  & \textbf{States} & \textbf{Time} \\
\hline
\multirow{3}{*}{single-pathing} 	&  MTTDIL	 & 5940 &  2.7 hr & 5972    & 0.54 & 13     & 137s\\ 
                        &  100\%SMTTDIL-R  & 37027 & 8 min     & 35671   & -3.6 & 2340   &  .31s\\
                        &   50\%SMTTDIL-R  & 10155 & 24 min      &  9048   & -11 &150 &.3  \\
                         &  100\%SMTTDIL+R & 2146215 & 7.1 hr & 2143599 & 0.12 & 2670   & 138s\\
                        &  50\%SMTTDIL+R   &  13975 & 5 hr & 14032    &  .4    & 150 & 137s\\ \hline
\multirow{3}{*}{multi-pathing}	&  MTTDIL 	  & 6968 &  3.6 hr & 7015    & 0.7 & 148    & 137s  \\ 
                        &   100\%SMTTDIL-R  & 37168 & 10 min & 36461 & -2  & 33495  & 3s \\ 
                        &    50\%SMTTDIL-R  & 14169 & 34 min     & 13171   & -7 & 5072& 0.4s  \\
                        &  100\%SMTTDIL+R   & 2578533 & 10 hr & 2581352 & 0.1  & 33495  & 160s  \\
                        &   50\%SMTTDIL+R  &  593232 & 9.9 hr & 597274  & .7  & 5072     & 138s \\ \hline
                                                                        
\end{tabular}
\end{center}
\caption{Results with Simulation and Hierarchical Decomposition for the systems of Fig \ref{fig5}); 
 Err: \% diff betw the two results, SMTTDIL: System MTTDIL}
\label{table5}
\end{table}
 From the
table, it is clear that 
the metric 50\%SMTTDIL+R 
expresses the advantage of using multi-pathing which other metrics do
not.
Also, repair has a significant impact on 50\%SMTTDIL+R.

From the results of Table \ref{table5} With multi-pathing,  we get 
17\% higher MTTDIL and 20\% higher System MTTDIL+R (100\%) compared to single-pathing at the cost of 14 SAS cables and 2 expanders. 
In this system,  if we replace 6 disk RAID5 with 24 disk RAID10 in each enclosure
then with multi-pathing we get 7\% higher MTTDIL and 2\% higher System MTTDIL+R (100\%) at the cost of 50 SAS cables and 2 expanders. 
Such calculations are important for cost-reliability trade-offs. In the first 
case multi-pathing seems to be a good option while in the second case it is not. 
The reason is that, in the latter case, the enclosure is less reliable (full enclosure MTTF = 11100 hr) and
RAID10 is more reliable than RAID5. 
For the 100\%System MTTDIL+R and 50\% System MTTDIL+R (in case of multi-pathing),  the contribution of subsystems 
(i.e. interconnect and disk subsystem)  is very high; this can be verified by sensitivity analysis. Computationally,  each of the
subsystems considered in isolation has a failure rate that is very
close to constant failure rate; we could not therefore reject the
hypothesis that each of the subsystems has a constant failure rate at even
20\% level of significance in the
Kolmogorov-Smirnov test. For other reliability metrics,  the contribution of subsystems
(i.e. interconnect and disk subsystem) is much less; the enclosure
 having an exponential failure
 distribution is the main contributor; this has been 
verified by sensitivity analysis. Hence  the results of hierarchical
decomposition and simulation is almost same for all the reliability measures we have calculated.  
\section{How to span a RAID group across enclosures ?}
\label{app2}
\subsection{Without correlated failure}
How should a storage system designer distribute 
the disks of a RAID group across enclosures to get the maximum reliability? For
a $n$ disk RAID5 group, it is best to distribute it across $n$
enclosures (because no single enclosure failure causes data inaccessibility of the whole RAID5 group) but this may not be cost effective or possible. 
Hence, one has to choose the best configuration among the sub-optimal solutions rather than having the luxury of choosing the optimal
solutions.  
Here we present a greedy algorithm 
to find the optimum configuration for any $f$ tolerant RAID group
given $n$ disks and $m$ enclosures (Algorithm \ref{algo}).
\begin{algorithm}[h]
\caption{Greedy algorithm for spanning}

\begin{algorithmic}
    
    \While {$n>0$}
     
     \If {$n \leqslant mf$} 

        \State Place $f$ disks at a time in each of the empty
        enclosures until number of disks left ($l$) is less than
        $f$. Place these $l$ disks in the next enclosure. 
          \State \textbf{break};          
     \Else
        
         \State $C \gets$ highest available capacity amongst
         remaining enclosures 
         \State $d \gets min(C,n) $ 
         \State Fill the chosen enclosure 
 with $d$ disks
         \State $n \gets n - d$ 
         \State $m \gets m-1$
     \EndIf
   
   \EndWhile
   
   \end{algorithmic}
 \label{algo}
\end{algorithm}

\begin{figure}[htp]
  \centering
  \subfigure[spanning increases reliability]{\label{fig4:a}\includegraphics[width=0.49\linewidth]{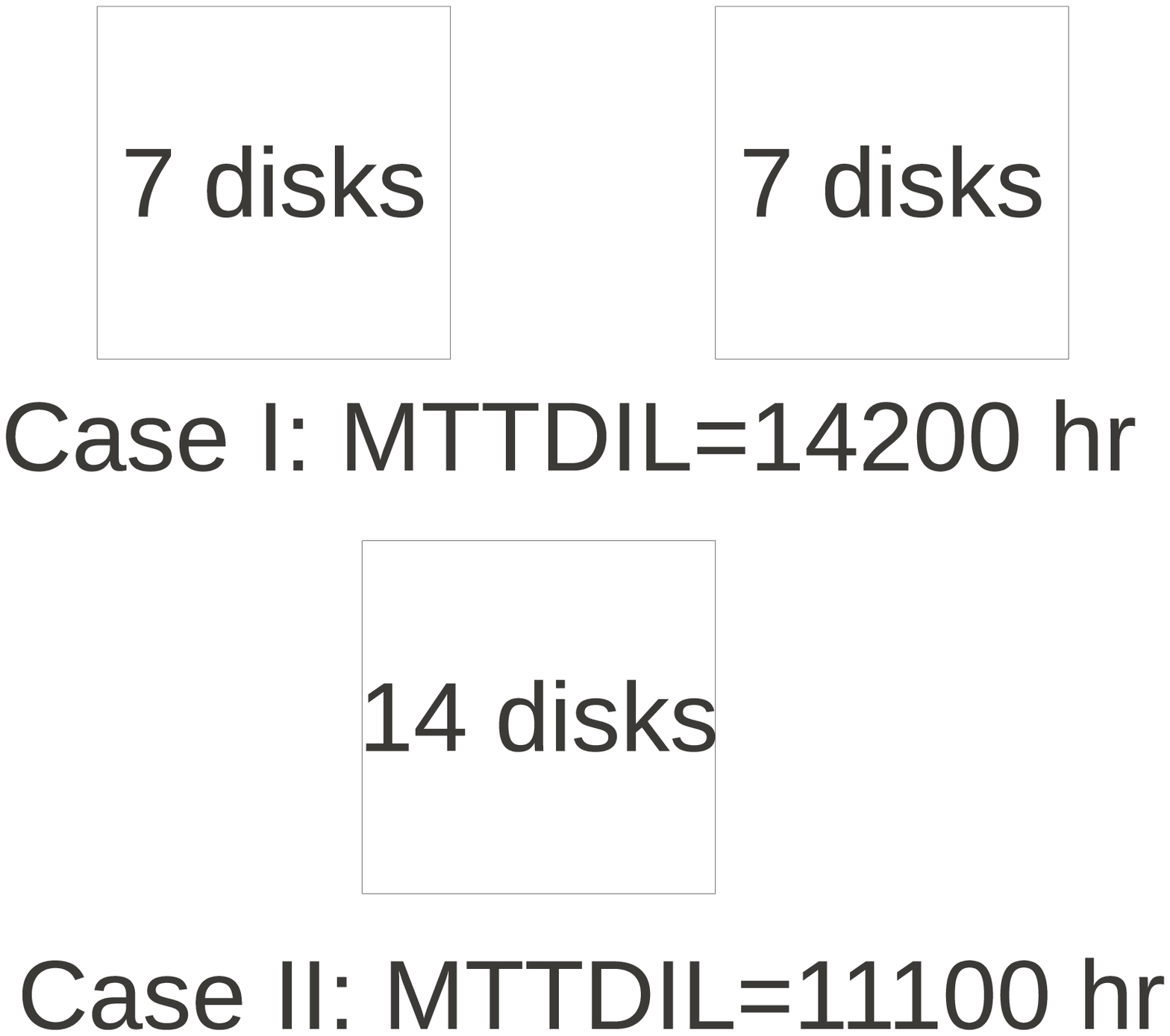}}\hfill
  \subfigure[spanning decreases reliability]{\label{fig4:b}\includegraphics[width=0.49\linewidth]{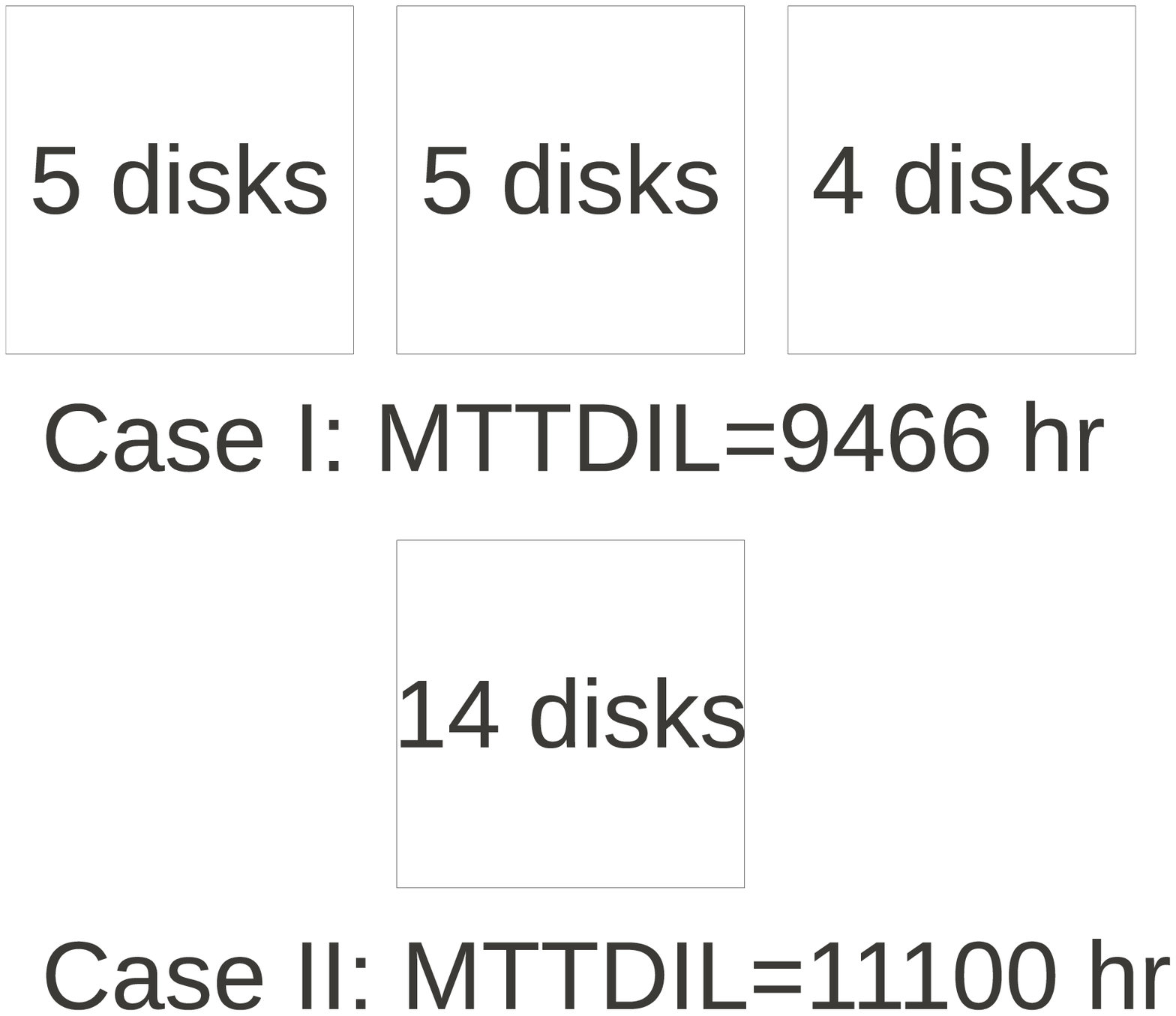}}
\caption{Enclosure failure rate $\propto$ no. of disks inside it}
\label{fig4}
\end{figure}

In Fig.\ref{fig4:a}, 
we need to design a 14 disk RAID5 group using 2 enclosures where the 
failure rate of an enclosure is as stated in Table \ref{table2}. 
Using algorithm \ref{algo}, 
the optimal configuration is to put all the disks 
in one enclosure. We obtain MTTDIL = 11100 hr.
Algorithm \ref{algo} assumes that all the enclosures 
have the same MTTF. 
However,  an enclosure is a shared component
whose failure rate depends on the number of disks present in it.
If we span the disks across 2 enclosures such that 
each of them contains less than or equal to 12 disks, we get MTTDIL = 14200 hr.
In Fig.\ref{fig4:b}, spanning a RAID group across 3
enclosures decreases reliability; here we used the PRISM simulator for
the computation.

\subsection{with correlated failure}
Spanning a RAID group across enclosures
Suppose,  a $n$ disk RAID5 group is formed inside an enclosure. Then
the rate at which data loss
occurs due to correlated failure is $^n\mathrm{C}_2 \lambda p$. Now,
if the $n$ disks are distributed across $m$ enclosures (where $m>1$
and, for simplicity, $n$ is a multiple of $m$ with each enclosure containing $n/m$ disks), then the 
rate at which data loss occurs due to correlated failure is $m[^{(n/m) }\mathrm{C}_2]\lambda p=\frac{(n((n/m) -1))  \lambda p}{2}$ 
that is less than $^n\mathrm{C}_2\lambda p$ for $m>1$. Hence  with respect to ``only'' correlated failures, spanning is a good option.

\begin{figure}[ht]
\centering
\includegraphics[height=1.5in]{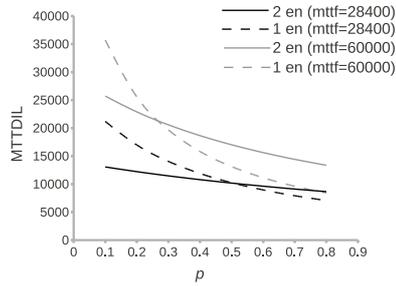}
\caption{MTTDIL (hr) for a RAID group of 8 disks using correlated disk
  failure model. en: enclosure 
}
\label{fig13}
\end{figure}

But,  whether spanning will increase the chance of overall ``data inaccessibility or data loss''  
will depend on enclosure failure rate also (Fig.\ref{fig13}). 
In Fig.\ref{fig13}, for $p$ = 0.4, spanning is beneficial when enclosure MTTF is 60000 hr.
but not useful if enclosure MTTF is 28400 hr. Similarly, for a given enclosure MTTF, say 60000 hr, spanning is beneficial 
when $p$ = 0.4 but not useful when $p$ = 0.2. 

\section{Memorylessness assumption of Markov Model and our solution:}
\label{app3}
Kevin Greenan  has raised several questions regarding suitability of Markov models as a tool to 
measure storage reliability \cite{Greenan}, as neither component wear-out nor rebuild progress can be modelled 
using a system level Markov model due to its memoryless property. The reason is that the notion of ``absolute'' time is present in a system 
level Markov model whereas ``relative time'' for each component is needed and simulation is the only solution. 

We propose a solution for this problem by considering failure and repair modes of each 
disk separately rather than considering a system level Markov model. Moreover, 
when we approximate Weibull repair and Weibull failure by summation of exponentials (i.e. by adding multiple states and transitions 
corresponding to a single failure/repair transition) then these states keep information regarding repair progress and age of a component 
respectively. Hence, our disk subsystem models using detailed disk models reduce the chance of 
loss of information due to memorylessness property significantly.

\begin{figure}[htp]
  \centering
  \subfigure[Simulation results \cite{Greenan}]{\label{greenan :a}\includegraphics[height=3 in]{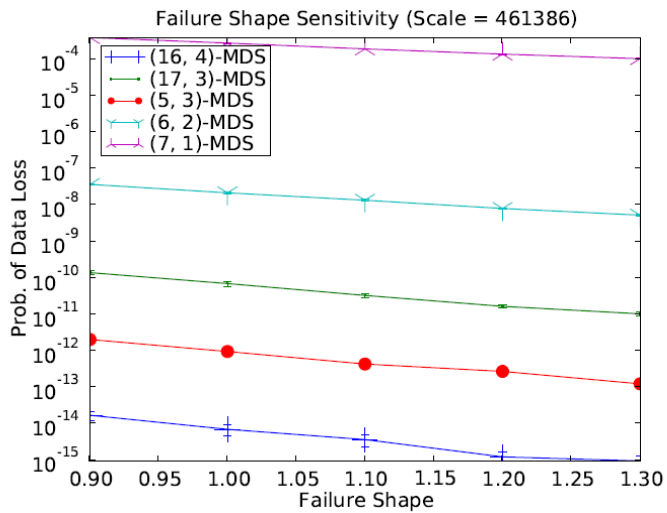}}\hfill
  \subfigure[PRISM results]{\label{greenan :b}\includegraphics[height=3 in]{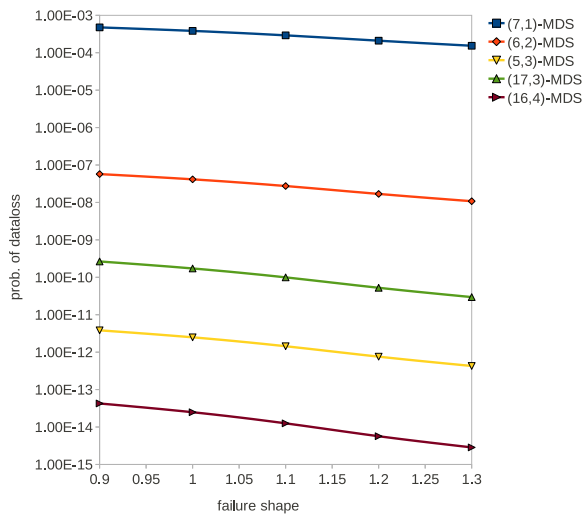}}
\caption{Comparison of PRISM results with Greenan's simulation results}
\label{greenan}
\end{figure}
To prove our claim, we modelled some disk subsystem configurations from Greenan's thesis \cite{Greenan} in PRISM and 
compared them with the Greenan's simulation results
(Fig. \ref{greenan}). We see that PRISM results are in reasonable
agreement with his simulation results.  

\end{document}